%%%%%%%%%
\input harvmac
%%% Figures
\input epsf
\ifx\epsfbox\UnDeFiNeD\message{(NO epsf.tex, FIGURES WILL BE
IGNORED)}
\def\figin#1{\vskip2in}% blank space instead
\else\message{(FIGURES WILL BE INCLUDED)}\def\figin#1{#1}\fi
\def\ifig#1#2#3{\xdef#1{fig.~\the\figno}
\goodbreak\midinsert\figin{\centerline{#3}}%
\smallskip\centerline{\vbox{\baselineskip12pt
\advance\hsize by -1truein\noindent\footnotefont{\bf
Fig.~\the\figno:} #2}}
\bigskip\endinsert\global\advance\figno by1}
%%%%%%%%%%%%%%%%%%%%%%%%%%%%%%%%%%%%%%%%%%%%%%%%%%%%%%%%%%%%%%%%

\def\footnotefont{\tenpoint}

\newwrite\ffile\global\newcount\figno \global\figno=1
\def\fig{fig.~\the\figno\nfig}
\def\nfig#1{\xdef#1{fig.~\the\figno}%
\writedef{#1\leftbracket fig.\noexpand~\the\figno}%
\ifnum\figno=1\immediate\openout\ffile=figs.tmp\fi\chardef\wfile=\ffile%
\immediate\write\ffile{\noexpand\medskip\noexpand\item{Fig.\
\the\figno. }
\reflabeL{#1\hskip.55in}\pctsign}\global\advance\figno
by1\findarg}
%\ionew

\overfullrule=0pt
\parindent 25pt
\tolerance=10000
%\sequentialequations
%\draftmode
%%%%%%%%% To write text on a figure

\def\at#1#2#3{\setbox0=\hbox{#3}\ht0=0pt\dp0=0pt \rlap{\kern#1\vbox
    to0pt{\kern-#2\box0\vss}}}

%%%%%%%%%

\def\NSNS{{$NS\otimes NS$}}
\def\RR{{$R\otimes R$}}
\def\ZZ {{\bf Z}}

\def\CC {{\bf C}}

\def\det{\hbox{\rm det}}

\def\G(#1){\Gamma(#1)}

\def\half{{\textstyle {1 \over 2}}}

\def\(#1#2){(\zeta_#1\cdot  \zeta_#2)} 
\def\ls{l_s}

\def\hn{\hat n}
\def\hm{\hat m}

\def\hlambda{\hat \lambda}
\def\hrho{\hat \rho}
\def\hsigma{\hat \sigma}
\def\hdelta{\hat \Delta}

\def\hDelta{\hat \Delta}
\def\hK{\hat K} 

\def\C|#1{{\cal #1}}
\def\calV{{\cal V}}
\def\calR{{\cal R}}
\def\calT{{\cal T}}
\def\GST{{\cal G}_{ST}}
\def\GTU{{\cal G}_{TU}}
\def\GUS{{\cal G}_{US}}
\def\calW{{\cal W}}
\def\calgst{\GST^s}
\def\calgtu{\GTU^s}
\def\calgus{\GUS^s}
\def\calw{{\cal W}^s}

\def\calF{{\cal F}}

%%%%%%%
\def\lr{\lref}
\lr\rfGreenVanhoveAnalytic{M.B. Green and P. Vanhove, {\sl The low-energy expansion of the one-loop type II superstring amplitude}, DAMTP/99-124, hep-th/9910056.}
\lr\rfGreenCargese{M.B. Green, {\sl Connections between M-theory and
    superstrings}, Proceedings of the 1997 Advanced Study Institute on
  Strings, Branes and Dualities, Cargese, France Nucl.Phys.Proc.Suppl. 68
  (1998) 242,  hep-th/9712195.} 
\lr\rfBachasLecture{C.P. Bachas, {\sl Lectures on D-branes}, lectures given in
  1997 at the Isaac Newton Institute, Cambridge, the Trieste Spring School on
  String Theory, and at the 31rst International Symposium Ahrenshoop in
  Buckow, hep-th/9806199.} 
\lr\rfDixonKapluLouis{L. Dixon, V. Kaplunovski and J. Louis, {\sl Moduli
    Dependence of String Loop Corrections to Gauge Couplings Constants},
  Nucl.Phys. {\bf B335} (1991) 649.}
\lr\rfBFKOV{C. Bachas, C. Fabre, E. Kiritsis, N.A. Obers and P. Vanhove, {\sl
    Heterotic / type I duality and D-brane instantons}, Nucl.Phys. {\bf B509}
    (1998) 33, hep-th/9707126.}
\lr\rfWittenVarious{E. Witten, {\sl String Theory Dynamics in Various
    Dimensions}, Nucl.Phys. {\bf B443} (1995) 85, hep-th/9503124.}
\lr\rfSchwarzPower{J.H. Schwarz, {\sl The Power of M-theory}, Phys.Lett. {\bf
    367B} (1996) 97, hep-th/9510086.}
\lr\rfAspinwall{P.S. Aspinwall, {\sl  Some Relationships Between Dualities in
    String Theory}, Nucl. Phys. Proc. Suppl. {\bf 46} (1996) 30,
    hep-th/9508154.}
\lr\rfCremmerJuliaScherk{E. Cremmer, B. Julia and J. Scherk, {\sl Supergravity
    Theory in Eleven-Dimensions}, Phys.Lett {\bf 76B} (1978) 409.}
\lr\rfBernDunbar{Z. Bern, L. Dixon, D.C. Dunbar, M. Perelstein and
    J.S. Rozowsky , {\sl On the Relationship between Yang-Mills Theory and
    Gravity and its Implication for Ultraviolet Divergences}, Nucl.Phys. {\bf
    B530} (1998) 401, hep-th/9802162;\hfill\break
 {\sl Perturbative Relationships Between QCD and Gravity and Some Implications},
  hep-th/9809163.}
\lr\rfGreenGutperleKwon{M.B. Green, M. Gutperle and H. Kwon, {\sl
    Sixteen-Fermion and Related Terms in M-theory on $T^2$}, Phys.Lett. {\bf
    B421} (1998) 149, hep-th/9710151.}
\lr\rfGreenGutperleKwontwo{M.B. Green, M. Gutperle and H. Kwon,
{\sl Light-cone quantum mechanics of the eleven-dimensional
superparticle}, JHEP 9908:012,1999, hep-th/9907155.}
\lr\rfGreenGutperleVanhove{M.B.Green, M. Gutperle and P. Vanhove, {\sl One-loop
    in Eleven Dimensions}, Phys.Lett. {\bf B409} (1997) 177, hep-th/9706175.}
\lr\rfGreenVanhoveMtheory{M.B. Green and P. Vanhove, {\sl D-instantons,
    Strings and M-theory}, Phys.Lett. {\bf 408B} (1997) 122, hep-th/9704145.}
\lr\rfRussoTseytlin{J.G. Russo and A.A. Tseytlin, {\sl One-loop four-graviton
    amplitude in eleven-dimensional supergravity}, Nucl.Phys. {\bf B508} (1997)
    245, hep-th/9707134.}
\lr\rfGreenSethi{M.B. Green and S. Sethi, {\sl Supersymmetry Constraint on
    Type IIB Supergravity}, Phys.Rev. {\bf D59} (1999) 046006,
    hep-th/9808061.}
\lr\rfBoris{B. Pioline, {\sl A note on non-perturbative $R^4$ couplings},
    Phys.Lett. {\bf B431} (1998) 73, hep-th/9804023.}
\lr\rfHerveAlex{H. Partouche and A. Kehagias, {\sl On the exact quartic
    effective action for the type IIB superstring}, Phys.Lett. {\bf B422}
    (1998) 109, hep-th/9710023; {\sl D-Instanton Corrections as (p,q)-String
    Effects and Non-Renormalization Theorems}, Int.J.Mod.Phys. {\bf A13}
    (1998) 5075, hep-th/9712164.}
\lr\rfDhokerPhongRevue{E. D'Hoker and D.H. Phong, {\sl The geometry of string
    perturbation theory}, Rev. Mod. Phys. {\bf 60} (1988) 917.}
\lr\rfGreenSchwarzWitten{M.B. Green, J.H. Schwarz and E. Witten, {\sl
    Superstring theory}, Cambridge University Press 1987.}
\lr\rfKawaiLewellenTye{H. Kawai, D.C. Lewellen and S.-H.H. Tye, {\sl A
    relation between tree amplitudes of closed and open strings},
    Nucl.Phys. {\bf B269} (1986) 1.}
\lr\rfGreenSchwarz{M.B.~Green and J.H.~Schwarz, {\sl Supersymmetrical Dual
    String Theory. (II). Vertices and Trees}, Nucl.Phys. {\bf B198} (1982)
  252.}
\lr\rfIengoZhu{R.~Iengo and C.~Zhu,
    {\sl Explicit modular invariant two-loop superstring amplitude
    relevant for  R**4,}  JHEP {\bf 06}  (1999) 011, hep-th/9905050.}

%%%%%%%%%%%%%%%%%%%%%%%%%%%%%%%%%%%%%%%%%%%%%%%%%%%%%%%%%%%%%%%%%%%
%%%%%%%%% title and abstract
%%%%%%%%%%%%%%%%%%%%%%%%%%%%%%%%%%%%%%%%%%%%%%%%%%%%%%%%%%%%%%%%%%%
\noblackbox
\baselineskip 14pt plus 2pt minus 2pt
\Title{\vbox{\baselineskip12pt
\hbox{hep-th/9910055}
\hbox{DAMTP/99-120}
\hbox{CERN-TH/99-283}
\hbox{SPHT-T99/100}
}}
{\vbox{
\centerline{Two  loops in eleven dimensions}
}}

\centerline{Michael B. Green, Hwang-h. Kwon and Pierre Vanhove}
\medskip
\centerline{DAMTP, Silver Street, Cambridge CB3 9EW, UK}
\centerline{\tt m.b.green,h.kwon,p.vanhove@damtp.cam.ac.uk}
\bigskip

%% abstract
\medskip
\centerline{{\bf Abstract}}
The two-loop Feynman diagram contribution to the four-graviton amplitude of
eleven-dimensional supergravity compactified on a two-torus, $\calT^2$, is
analyzed in detail. The Schwinger parameter integrations are re-expressed as
integration over the moduli space of a second torus, $\hat \calT^2$, which
enables the leading low-momentum contribution to be evaluated in terms of maps
of $\hat \calT^2$ into $\calT^2$.  The ultraviolet divergences associated with
boundaries of moduli space are regularized in a manner that is consistent with
the expected duality symmetries of string theory.  This leads to an exact
expression for terms of order $D^4\, \calR^4$ in the effective M theory action
(where $\calR^4$ denotes a contraction of four Weyl tensors), thereby
extending earlier results for the $\calR^4$ term that were based on the
one-loop eleven-dimensional amplitude.  Precise agreement is found with terms
in type IIA and IIB superstring theory that arise from the low energy
expansion of the tree-level and one-loop string amplitudes and predictions are
made for the coefficients of certain two-loop string theory terms as well as
for an infinite set of D-instanton contributions.  The contribution at the
next order in the derivative expansion, $D^6\,\calR^4$, is problematic, which
may indicate that it mixes with higher-loop effects in eleven-dimensional
supergravity.

%%%%%%%%%%%%%%%%%%%%%%%%%%%%%%%%%%%%%%%%%%%%%%%%%%%%%%%%%%%%%%%%%%%
\noblackbox
\baselineskip 14pt plus 2pt minus 2pt

\Date{PACS: 04.65.+e; 04.50.+h}

%%%%%%%%%%%%%%%%%%%%%%%%%%%%%%%%%%%%%%%%%%%%%%%%%%%%%%%%%%%%%%%%%%%
\newsec{Introduction}

This paper continues the study of the interconnections between
quantum supergravity in eleven dimensions  \refs{\rfCremmerJuliaScherk}  compactified on
$\calT^2$
and properties of perturbative and nonperturbative string theory
\refs{\rfAspinwall,\rfSchwarzPower}.
In earlier papers
\refs{\rfGreenGutperleVanhove,\rfGreenGutperleKwon,\rfGreenGutperleKwontwo}
it was shown that
the one-loop
diagrams of eleven-dimensional supergravity
that contribute to certain special  amplitudes reproduce terms
in the effective type II superstring actions that may be described
by integrals over sixteen Grassmann components, which is half the
dimension of the type II superspace.  These terms include the
$\calR^4$ term, which is a specific contraction of four Weyl tensors
 that arises from the leading behaviour in the low energy expansion
 of the four-graviton amplitude.

The main  objective of this paper is to extend
this analysis to the evaluation of higher-derivative
 terms in the effective action by considering the low
energy expansion of the two-loop contribution to
eleven-dimensional supergravity compactified on $\calT^2$.  This
seemingly awesome calculation is greatly facilitated by the
observation in \refs{\rfBernDunbar} that the two-loop amplitude has a
surprisingly simple expression as a kinematic factor multiplying
a subset of the two-loop amplitudes of $\varphi^3$ {\it scalar} field
theory.  This is a generalization of the   well-known  structure of  the one-loop amplitude.

Eleven-dimensional supergravity is only the long wavelength
approximation to M theory and does not by itself define the short
distance physics that is necessary for a consistent quantum theory.
 This is evident from the fact that the quantum theory has
non-renormalizable  ultraviolet behaviour that can only be consistently
interpreted  with additional microscopic input that is not contained in
the supergravity theory but should be built into a
  detailed microscopic theory, such as the matrix model.
 However, it was seen in \refs{\rfGreenGutperleVanhove} that if some  mild
 extra information is fed in from string theory the regularized value of the
 one-loop divergence in the four-graviton scattering amplitude
 is uniquely specified.  This mild information is the fact
 that the type IIA and IIB superstring theories have identical
 one-loop four-graviton amplitudes.  Similar statements hold for other interactions
 of the same dimension that are related to the four-graviton interaction by
 supersymmetry \refs{\rfGreenGutperleKwon}.
 We will see that requiring the various string duality
  symmetries to hold  will also  severely restrict
  the form of  special higher-dimension interactions that arise
  at two loops in eleven-dimensional supergravity and contribute
   higher-derivative  terms in the effective action.

In section 2 we will give a schematic overview  of the  loop
amplitudes of eleven-dimensional supergravity compactified on a circle
and on a two-dimensional
torus, and their correspondence with terms in the string theory
effective action.  The purpose of this section is to show how
 simple dimensional arguments can hint at connections between these quantum
 loop amplitudes and the structure of higher-order terms
 in the effective action of type II string theory in nine and ten dimensions.
An important point to be discussed at the end of section 2 is
that the  four-graviton amplitudes in
the ten-dimensional type IIA and type IIB theories can be shown to be
{\it equal} up
 to two loops, even though the two-loop amplitudes are notoriously difficult to
evaluate in closed form.   This rather non-obvious consequence of
supersymmetry
follows from careful consideration of the effect of the insertion of
world-sheet supermoduli.

In order to compare our results obtained from one and two-loop
diagrams of
eleven-dimensional supergravity on $\calT^2$ with the corresponding
string theory results, we include an appendix which contains a brief  review of the
expansion of the four-graviton
tree-level and one-loop string theory amplitudes
in a series of derivatives.

Section 3 will  review the detailed calculation of the one-loop
four-graviton amplitude compactified on $\calT^2$ and its
contributions to the effective M-theory action, developing the
arguments  in \refs{\rfRussoTseytlin,\rfGreenCargese} concerning
the momentum dependence. The lowest order term in the momentum
expansion  determines the interaction of the form \refs{\rfGreenVanhoveMtheory},
\eqn\lowes{\int d^9
x\sqrt{-G^{(9)}} \calV\, \calR^4\, \left({4\pi\over 3}\Lambda^3l_{11}^3 +
  \calV^{-{3\over 2}}f_1(\Omega,\bar\Omega)
\right),}
 where
$G^{(9)}$ is the nine-dimensional M-theory metric, $\calV$ is the
dimensionless volume of $\calT^2$, $\Omega=\Omega_1+i\Omega_2$ is
its complex structure and $f_1(\Omega,\bar\Omega)$ is a
modular-function invariant. When translated into the
nine-dimensional type IIB string theory parameters the complex
structure is identified with the complex coupling constant
 where $\Omega_1$ is the Ramond--Ramond (\RR)
scalar field and $\Omega_2=e^{-\phi^B}$, with $\phi^B$ the type IIB
dilaton.  The function $f_1(\Omega,\bar\Omega)$ has a
large-$\Omega_2$ expansion that begins with two power-behaved
terms.  These are interpreted in string theory as terms that arise from
the tree-level string amplitude and from the one-loop string
amplitude.  The remainder of $f_1(\Omega,\bar\Omega)$ consists of
an infinite sequence of exponentially suppressed contributions of
the form $e^{-2\pi (|K|\Omega_2 -i K\Omega_1)}$ which correspond to
D-instanton contributions.  The one-loop ultraviolet divergence is cubic in the loop
momentum and has been cut off in \lowes\ at a momentum scale $\Lambda$
measured in units of $l_{11}^{-1}$, where $l_{11}$ is the eleven-dimensional
Planck length.  It was shown in \refs{\rfGreenVanhoveMtheory} that in order for
\lowes\ to be consistent with the equality of the
 one-loop four-graviton amplitudes in the   IIA and IIB string theories
 the cut-off must be set to  the
 value $(\Lambda l_{11})^3 = \pi/2$.  Alternatively,  a local $\calR^4$ counterterm should be added to the
one-loop action  with a coefficient chosen to cancel the $\Lambda$ dependence
and give the appropriate finite value.

The one-loop amplitude compactified on $\calT^2$
also contains an infinite set of higher-derivative terms.
Among these are the non-analytic terms containing the
nine-dimensional massless threshold
singularities  implied by unitarity which have the symbolic form
$(-l_s^2 s)^{1/2}$ (where $s$
represents any of the Mandelstam invariants\foot{We use lower-case
letters $s,t,u$ to denote the Mandelstam invariants in the ten-dimensional
theory in the string frame and
upper-case letters $S,T,U$ for the corresponding invariants in the
eleven-dimensional theory.}).  After subtracting
this term the loop amplitude can be expanded in a  series of
powers of the momenta corresponding to higher derivative
terms in the effective action
\refs{\rfRussoTseytlin,\rfGreenCargese}.  These higher-derivative
terms translate into terms in the IIA and IIB string theory effective actions
that have a dependence on the coupling constant that implies that
they should be identified with  multi-loop string theory effects.
Among these  terms are contributions of order $s^2\calR^4$  that have the
dilaton dependence of string theory one-loop and two-loop
contributions.  These terms  apparently violate the equality of  IIA
and IIB  four-graviton amplitudes at two string loops.  However, we
will see that the expected equality   is restored when  two-loop
supergravity effects are added.

Section 4 will be concerned with  the two-loop supergravity four-graviton
amplitude compactified on $\calT^2$,  making use of  its expression in terms
of  scalar field theory \refs{\rfBernDunbar}.  An important feature of
the two-loop and higher-loop contributions is that they  have overall
kinematic factors of the form $D^4\calR^4$, so that they do not give extra
contributions to  the one-loop $\calR^4$ term.\foot{This symbolic notation
  indicates a term in which there are four (covariant) derivatives and four
  factors of the Riemann curvature.  The precise pattern of index contractions
  will be specified later.}     However, it is not known if the $D^4\calR^4$
and $D^6\calR^4$ terms, which get contributions from both one and two loops in
eleven dimensions, are protected from corrections arising from   higher-loop
diagrams.  In a sense, the results of this paper indicate that the
$D^4\calR^4$ terms are completely accounted for by the two-loop  contributions
and should therefore not receive higher-order corrections.

We will be interested in the expression for the two-loop amplitude compactified on
 $\calT^2$ so that each loop is associated with an independent
 nine-dimensional momentum
 integral and a sum over two Kaluza--Klein momentum components.
After  performing the integration over the continuous loop momenta the
leading term in the low energy expansion of the
two-loop supergravity amplitude of $D^4\calR^4$ will be
expressed as an  integral
over three Schwinger parameters and a sum over the Kaluza--Klein charges.  This  needs to
be regularized in order to suppress the
ultraviolet divergences which are of two kinds.
 One of these is the two-loop primitive divergence while the second
comprises the three independent sub-divergences that come from the
divergences of one-loop sub-diagrams.

In order to describe these divergences in a systematic manner we first
 perform Poisson resummations over the Kaluza--Klein momenta to rewrite
 the amplitude as a  sum over the windings of the internal lines around $\calT^2$
  as well as an integral over  three loop parameters.    The leading divergence
  arises, as expected, in the sector of zero winding number while the one-loop
  sub-divergences arise
  in sectors in which a subset of winding numbers vanish.
 In order to analyze these sub-divergences  we have found it very
 helpful to make use of a hidden $SL(2,\ZZ)$ symmetry of the two-loop
 supergravity integrand.
This is made explicit by redefining the three loop  integration
 variables to be the volume and complex structure of a
second two-torus, $\hat{\calT^2}$.  The ultraviolet divergences  are regularized
in a natural manner that respects the $SL(2,\ZZ)$ symmetry
 by introducing a cutoff at the boundaries of the moduli space of this torus.
The evaluation of the loop amplitude then involves
mappings of $\hat{\calT^2}$ into $\calT^2$.

  In this manner we will be
able to evaluate contributions to the effective action that have the form of a
prefactor,  which is a function of the moduli, multiplying $D^4\calR^4$.
 There is a
finite (cutoff-independent) contribution to this prefactor
that is independent of the string coupling and is interpreted
as a string one-loop contribution.   The
 dependence on the complex structure of the torus is encoded in a contribution to
  the prefactor that is
 again a modular invariant non-holomorphic Eisenstein series.  This enters in the
 sectors that have one-loop sub-divergences proportional to $\Lambda^3$, where
 $\Lambda$ is a momentum  cutoff.
 These sub-divergences are cancelled by additional  one-loop
  four-graviton diagrams in which  the one-loop $\calR^4$
   counterterm  (and its supersymmetric partners) is  inserted as one of the vertices.
When translated into string theory coordinates the
renormalized value of this prefactor  contains
    equal tree-level, one-loop and two-loop
     perturbative contributions to the type IIA and IIB string theory
 four-graviton amplitudes.
 The agreement between these IIA and IIB perturbative terms
follows from detailed comparison between  the one-loop expressions of section 3 and the
two-loop expressions of section 4.  The coefficients of these terms
are also in precise agreement  with the corresponding
terms in the expansion of the string tree and one loop amplitudes
given in  the appendix.

 We will also argue on the basis of
string dualities that the leading two-loop divergence
can make no contribution to the $D^4\calR^4$  interaction, which means that its
renormalized value must be set equal to zero.
However, it can contribute  to the $D^6 \calR^4$ interaction at
string tree level.  The analysis of the eleven-dimensional
two-loop contribution to this interaction indeed appears to be a
mess.   This suggests that this  interaction may also receive
contributions from higher-loop effects in eleven-dimensional
supergravity.

Section 5 contains a summary and some concluding comments.

%%%%%%%%%%%%%%%%%%%%%%%%%%%%%%%%%%%%%%%%%%%%%%%%%%%%%%%%%%%%%%%%%%%%%%%%%%%%
\newsec{Higher order terms in eleven-dimensions}

The  derivative expansion of the M-theory action for the eleven-dimensional
theory compactified on $\calT^2$ starts with the classical
Einstein-Hilbert term,
\eqn\ehilb{
\C|S_{EH} = {1\over 2 \kappa_{11}^2 } \int d^{11}x \sqrt{-G^{(11)}} \ R\ ,
}
where  $2\kappa_{11}^2=(2\pi)^8l_{11}^9$ and $l_{11}$
is the eleven-dimensional  Planck
length.\foot{With this convention the value of the tension
  of the fundamental string is equal to the tension of the
 M2-brane wrapped on a circle of radius
  $2\pi R_{11}$, i.e.  $T_F=2\pi R_{11}l_{11} (2\pi^2/\kappa_{11}^2)^{1/3}$
\refs{\rfSchwarzPower,\rfBachasLecture}.} There is no
 coupling constant that can be tuned to a small value
in the eleven-dimensional theory so there is no meaningful
perturbative expansion.  Therefore, we will only be able to make sense of
`protected' quantities that receive only a finite number of perturbative
contributions.   The dimensional ultraviolet cutoff is
determined in units of the eleven-dimensional Planck scale,
$l_{11}$.  Upon compactification it will
often be convenient to change to the string theory parameters,
which are given in units of the string scale,  $l_s$.
Compactification on a circle of radius $R_{11}$
 gives rise to the type IIA string theory
where  the string coupling constant,
$g^A=e^{\phi^A}$ (where $\phi^A$ is the IIA dilaton),
is given  by $l_{11} = (g^A)^{1/3} \ls$ and
$R_{11}^3=e^{2\phi^A} =
(g^A)^2$. Masses are measured with the metric \refs{\rfWittenVarious}
\eqn\eMetric{
ds^2=G^{(11)}_{MN}dx^Mdx^N=
 {l_{11}^2\over l_s^2R_{11}}\; g_{\mu \nu} dx^\mu dx^\nu +
R_{11}^2 l_{11}^2 (dx^{11} - C_\mu dx^\mu )^2,
}
where $g_{\mu \nu}$ is the string frame metric.
Since the compactification radius
$R_{11}$ depends on the string coupling constant the Kaluza-Klein modes are
mapped to the massless fundamental string states  and
the non-perturbative D0-brane states.
When expressed in terms of the type IIA string theory parameters the
compactified classical action becomes
\eqn\iiaclass{S_{EH} = {1\over 2 \kappa_{10}^2} \int d^{10} x\, \sqrt{- g} \,
e^{-2\phi^A}\, R,}
where  $2\kappa_{10}^2=(2\pi)^7 l_s^8$ and $l_s$ is the string length
scale.\foot{In this convention the fundamental string tension is
related to the string scale by
$T_F^2 = \pi (2\pi l_s)^4/\kappa_{10}^2$.}

More generally, we will be concerned with the compactification of
eleven-dimensional supergravity on $\calT^2$.  The dictionary
that relates $\calV$ and $\Omega$ to the nine-dimensional type IIA and type IIB
string theory parameters is  \refs{\rfAspinwall,\rfSchwarzPower},
\eqn\usdico{\eqalign{
\calV=  &R_{10}R_{11}= \exp\left({1\over 3}\phi^B\right)
r_B^{-{4\over 3}},\qquad   r_B={1\over R_{10} \sqrt{R_{11}}} =
r_A^{-1},
\cr
&  \Omega_1 = C^{(0)} = C^{(1)}_9, \qquad  \Omega_2= {R_{10}\over
 R_{11}}=\exp\left(-\phi^B\right)=r_A\, \exp\left(-\phi^A\right),\cr}}
where $r_A$ and $r_B$ are the   dimensionless radii of the tenth
dimension as measured in the IIA and IIB string frames,
respectively.  The one-form $C^{(1)}$ and the zero-form $C^{(0)}$
are the respective \RR\ potentials and $\phi^A$, $\phi^B$ are the
IIA and IIB dilatons.

The higher order corrections to the four-graviton interaction in the
 M-theory effective action compactified on
  $\calT^2$ can be schematically represented by the
expression,
\eqn\eMCorrections{
\C|S_4\sim  {1\over l_{11}} \; \int d^9x \sqrt{-G^{(9)}} \calV \;
  h(\calV,\Omega;l_{11}^2\partial^2) \calR^4\ ,
}
where the  expansion of the
function $h$ summarizes general features of the higher order
corrections to the action.   The lowest-order contribution of this type
is the   $\calR^4$ term
\refs{\rfGreenVanhoveMtheory,\rfGreenSethi,\rfBoris,\rfHerveAlex,\rfGreenGutperleVanhove},
which denotes the
familiar contraction between four Weyl tensors,
\eqn\erfdef{\calR^4\sim t^{\mu _1\dots \mu _8}
  t^{\nu_1\dots \nu_8} R^{\mu _1 \nu_1 \mu _2\nu _2} \dots
  R^{\mu _7\nu_7\mu _8\nu_8},}
  where the tensor $t^{\mu _1\dots \mu _8}$ ($\mu _r = 0,1,\cdots,9$)
  is defined in \refs{\rfGreenSchwarz}.
In the following we will be evaluating the scattering
 amplitude for four on-shell gravitons  that contributes
  to effective interactions of this type.
 Instead of specifying the precise normalization constant in \erfdef,
 it will therefore be  more useful to define the linearized version of this
 interaction in momentum space, which is given by
\eqn\useam{ \hK =
t^{\mu _1\dots \mu _8} t^{\nu_1\dots \nu_8} \prod_{r=1}^4 \zeta^{(r)}_{\mu _r\nu_r}
k^{(r)}_{\mu _{r+4}} k^{(r)}_{\nu_{r+4}},}
where  $\zeta^{(r)}_{\mu _r\nu_r}$ ($r=1,2,3,4$) are the polarization
vectors for the gravitons with  momenta $k^{(r)}_{\mu _r}$
  satisfying the  conditions
$k^{(r)\, 2}=0$  and  $\sum_{r=1}^4
k_r =0$.

In writing the effective action in the form \eMCorrections\ it is
necessary to first subtract the nonlocal threshold terms that arise from
integration over the massless  intermediate states in the loop
amplitudes.  In the nine-dimensional compactification to be
considered later these thresholds generate square root branch
points of the form $(-s)^{1/2}$.  Having subtracted this behaviour the
amplitude has a power series expansion in powers of $s$, $t$ and
$u$.  This translates into an expansion of the function $h$ in
powers of $\partial^2$, beginning with terms that we will write
symbolically $D^4\, \calR^4$, which have the linearized form,
\eqn\linad{\partial^4 \calR^4 \sim   t^{\mu _1\dots \mu _8}
  t^{\nu_1\dots \nu_8}\, (\partial_{\mu _2}\partial_{\nu_2}
  h_{\mu _1\nu_1})(\partial_{\mu _4}\partial_{\nu_4}
  h_{\mu _3\nu_3})\partial^4 \left((\partial_{\mu _6}\partial_{\nu_6}
  h_{\mu _5\nu_5}) (\partial_{\mu _8}\partial_{\nu_8}
  h_{\mu _7\nu_7})\right).}
The precise normalization of this term will be relevant later when
its contribution to the four-graviton amplitude will be discussed.  In
that case we will define $D^4\,\calR^4$ in such a manner that it gives
  a four-graviton contact term that is equal to
\eqn\defcals{(S^2+T^2+U^2)\, \hK.}
The possible term of order $\partial^2$
vanishes by use of the equations of motion (the mass shell
condition $S+T+U=0$).  There are  expected to be higher order
non-analytic terms of the form $S^3 (-S)^{1/2}$ which will also
need to be subtracted before powers of $S^4$ and higher can be
considered.  However, the considerations of this paper will cover
only the terms of order $S^2\calR^4$ (together with a few
comments about terms of order $S^3\calR^4$) so the higher-order
thresholds will not be relevant.

%%%%%%%%%%%%%%%%%%%%%%%%%%%%%
\ifig\fone{(a) The scalar field theory
 one-loop diagram contributing to the four-graviton amplitude
of compactified eleven-dimensional supergravity.  (b)  The one-loop $\calR^4$
 counterterm that cancels the cubic ultraviolet divergence.}
{\epsfbox{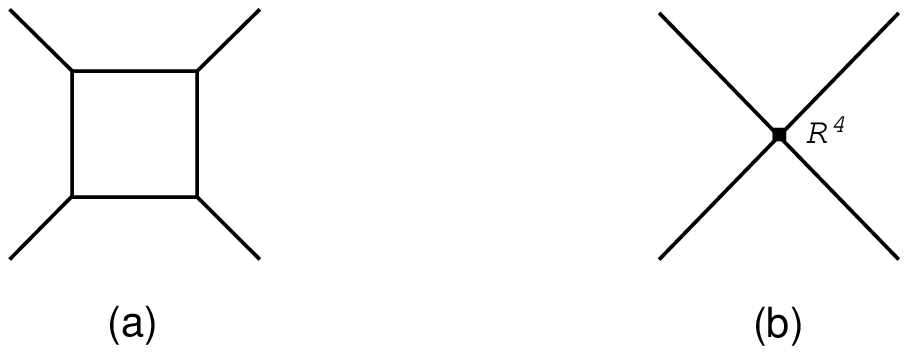}}
%%%%

\subsec{One-loop contributions}

Some of the systematics of the correspondence between the loop
calculations of eleven-dimensional supergravity compactified on a
circle or on a two-torus can be understood from dimensional
arguments.  For example, the one-loop four-graviton diagram of \fone\
has dimension $(momentum)^{11}$ but it actually only diverges
cubically with momentum.   This follows from the fact that an
overall factor of the linearized approximation to $\calR^4$
factors out of the amplitude and this prefactor
contains eight powers of the external momenta.  After accounting for this
prefactor the dynamical part of the loop calculation is identical to the box diagram
of $\varphi^3$ field theory.  Importantly,  no other diagrams
contribute.  In particular, there are no diagrams with vertices corresponding to
 gravitational contact interactions.  This simplification  is a
 very special feature of the four-graviton amplitude and other
 related processes that are protected by supersymmetry
 \refs{\rfGreenGutperleKwontwo}.

The box  diagram can
be expressed as a sum over the windings of the world-line of the particle
circulating in the loop, which gives an expression that is the
sum of integer winding numbers around the circle or the
two-torus.  The term with zero winding number, which is
ultraviolet divergent, does not depend on the geometry of the
torus.   This divergence will be regulated by introducing a cutoff,
$\Lambda^{-2}$, on the Schwinger parameter conjugate to the loop momentum
which suppresses the ultraviolet domain.  This
gives a contribution to  the amplitude  proportional to $\Lambda^3$.  The
dependence on the cutoff can be cancelled by adding a local $\calR^4$ counterterm
to the action. The sum over non-zero windings gives a finite
contribution which is necessarily proportional to $R_{11}^{-3}$
which has the dimensions $(momentum)^3$  where, for  simplicity, we are
considering compactification on a circle.  Comparing this
to the expected result in the type IIA theory, in which $R_{11}^3 =
e^{2\phi^A}$, we see that the
finite term is interpreted as a string  tree-level effect while the
regularized term (which is independent of $e^{\phi^A}$) is a
string one-loop effect.  Compactification of the loop on a two-torus
gives a  dependence on the modulus of the torus as well as its
volume.  In the limit of zero volume the ultraviolet divergent zero
winding number
term vanishes and the finite sum over non-zero windings gives the finite
result that corresponds to the type IIB string theory.

As will be explained in sub-section 2.3  the
four-graviton amplitudes of the IIA and IIB string perturbation
expansions  are
equal up to and including two loops.
This is not an automatic property of the  eleven-dimensional field theory
 calculation but it is true if
the coefficient of the counterterm is chosen to have an appropriate value.
Furthermore, this is the
same value that is required by supersymmetry (based on an
indirect argument given in \refs{\rfGreenVanhoveMtheory}).

In the next section we will consider terms of higher order in the
derivative expansion that come from the expansion of the one-loop
supergravity amplitude  in powers of $S,T$ and $U$.  When compactified on a circle
this gives terms in the effective action
of the symbolic form $R_{11}^{3n}\, D^{2n}\, \calR^4$,
which contribute  to the $n$-loop string action since $R_{11}^3 = e^{2\phi^A}$.

%\break
\subsec{Two-loop contribution}

 %%%%%%%%%%%%%%%%%%%%%%%%%%%%%
\ifig\ftwo{(a) A scalar field theory two-loop diagram that contributes
 to four-graviton scattering.   (b)  One-loop diagrams
in which one vertex is the $\calR^4$ counterterm cancel
the   sub-divergences of the two-loop diagrams.   (c)  A two-loop counterterm
proportional to $S^2\, \calR^4$.}
{\epsfbox{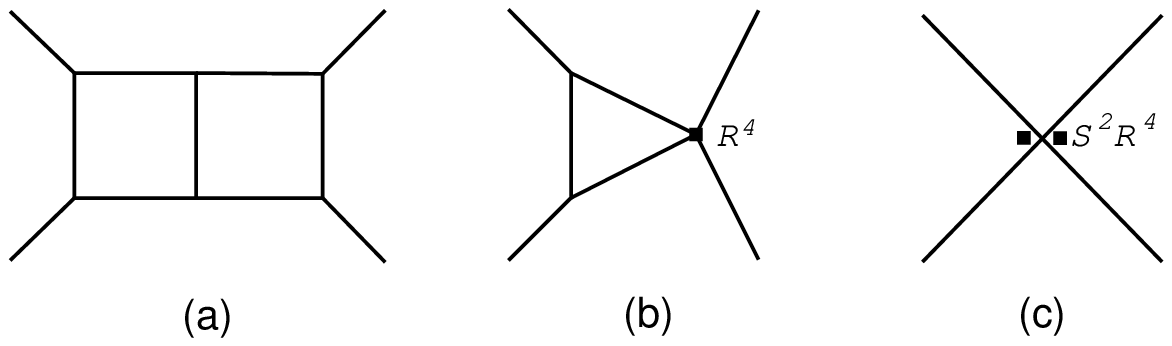}}
%%%%
New primitive divergences arise at each order in perturbation theory.
For example, the two-loop Feynman diagrams contributing to the
four-graviton amplitude in eleven-dimensional supergravity have the
 superficial degree of divergence $\Lambda^{20}$.
However,  the amplitude has an expansion in powers of  derivatives
acting on  $\calR^4$  so   there are at
 least eight powers of the external momenta, reducing the naive
divergence to $\Lambda^{12}$, or less (depending on the number of
derivatives).  According to \refs{\rfBernDunbar} at two loops there is also an
additional factor of $S^2$ (or $T^2$ or $U^2$) so that the naive two-loop
divergence is $\Lambda^8$, which is the same as that of $\varphi^3$
scalar field theory.
More generally, at $n$ loops there is a new  primitive divergences of the form
$\Lambda^{9n-10} S^2 \calR^4$.  From the work of \refs{\rfBernDunbar}
 it is not yet clear  whether extra overall
powers of $S$, $T$ and  $U$
arise beyond two loops which would reduce the naive degree of divergence
still further (although it seems unlikely that there will be a simple
expression for higher loops in terms of $\varphi^3$ field theory).
These  ultraviolet divergences  come from
the sector in which all winding numbers vanish and are   independent of
the geometry of the compactified dimensions. Their
cutoff dependence  can therefore be
subtracted by the inclusion of local counterterms proportional to
powers of derivatives
acting on four powers of the curvature (as in \ftwo(c)).  In
\refs{\rfBernDunbar} the two-loop amplitude was evaluated by
dimensional regularization, which picks out the logarithmically
divergent term.  This  arises from the finite  part of a term of the  (symbolic)
form $S^{6-2\epsilon} \, \calR^4\,/\epsilon$ in $11-\epsilon$ dimensions.
Likewise, the diagram will contribute non-analytic threshold
terms at order $S^6$.  Dimensional regularization discards the power
divergences that have lower powers of $S$, which are precisely the terms
 we are interested in this paper.

 In translating to string theory we must
use the relations between the string theory Mandelstam invariants,
$s$, $t$ and $u$,
and those of eleven-dimensional supergravity,
\eqn\relman{s = S{l_{11}^2\over l_s^2R_{11}}, \qquad t=T{l_{11}^2\over
l_s^2R_{11}}, \qquad u=U{l_{11}^2\over l_s^2R_{11}},}
where the presence of the inverse powers of $R_{11}$ results from the inverse
metric in the definition of the invariants ($S=- G^{\mu \nu} (k_1+k_2)_\mu 
(k_1+k_2)_\nu$, $T= -G^{\mu \nu} (k_1+k_4)_\mu 
(k_1+k_4)_\nu$, $U=- G^{\mu \nu} (k_2+k_4)_\mu 
(k_2+k_4)_\nu$).

As in the case of one-loop diagrams the effects of the internal propagators
winding around the compact direction(s) leads to dependence on the geometry
of the compact dimensions.  In this case these effects arise both in finite
terms as well as in terms that contain sub-divergences.  For
example, fig.2(a) shows an example of a two-loop
supergravity diagram   which has
dimension $(momentum)^{20}$.
 After accounting for the twelve powers of the
external momenta in the overall $S^2\, \calR^4$ factor eight powers of momenta remain
that must be replaced either by powers of $\Lambda$ or
appropriate powers of the dimensional parameters, $(R_{11})^{-1}$ and
$s=S/R_{11}$, $t=T/R_{11}$, or $u=U/R_{11}$.
When compactified on a circle of radius $R_{11}$ this
will contribute to the string tree level amplitude if it is
 proportional to $(R_{11})^{-3}$.    There are therefore two possible kinds
 of term that contribute at tree level, namely, terms of the form,
 \eqn\oneterm{{K_1\over R_{11}^3}\, \Lambda^3\, l_{11}^4\left({S^2 +T^2 +
U^2\over R_{11}^2 }\right) = g_A^{-2}\, \Lambda^3\, l_s^4(s^2+t^2+u^2),}
 and
\eqn\twoterm{{K_2\over R_{11}^3}\,  l_{11}^6\left( {S^3 +T^3 + U^3\over
 R_{11}^3 }\right) = g_A^{-2} \, l_s^6(s^3+t^3+u^3),}
 where $K_1$ and $K_2$ are constants.
 The second of these terms does not depend on the cutoff
 and   is a finite contribution whereas the first term results from  the one-loop
 sub-divergences.

 These  sub-divergences are cancelled by including the one-loop diagram of
 \ftwo(b), in which the vertex indicated by the dot is  the local $\calR^4$
 counterterm that was extracted from the one-loop diagram and has a coefficient
 that depends on the cutoff.  Since
the particles circulating in the loop include all components of the supermultiplet,
the supersymmetric partners of the $\calR^4$ vertices are also involved.  These
couple the two external gravitons to two internal third-rank antisymmetric tensors,
or two gravitini, in addition to two internal gravitons.  In practice,
this complication will be avoided since  we will find that the consistency of the
renormalization procedure requires \ftwo(b) to be
given in terms of scalar field theory in the same manner as the other one-loop
 and two-loop diagrams.  This makes the diagram very easy to evaluate.
 Its dimension of  $(momentum)^{17}$  is
accounted for by the cutoff-independent
factor $R_{11}^{-3}\,(S/R_{11})^2\,\calR^4$ that has the same form as \oneterm.
The sum of this diagram and \oneterm\ should give a specific overall coefficient
 that is independent of the cutoff.
In fact, we  will see from the explicit calculations
in section 4 that the coefficient of the $D^4\, \calR^4$ term is
 proportional,  in the type IIB limit, to
$E_{5/2}(\Omega,\bar\Omega)$, which is the natural modular invariant completion of
the tree-level term \oneterm. As anticipated,  the overall  coefficient is
precisely determined by requiring that the type IIA and type
IIB string loop amplitudes are equal (up to two loops).

 The status of the $S^3 \, \calR^4$ term
 \twoterm\ will not be resolved in this paper.  It seems likely that a
 complete understanding will have to take into account higher-loop
 supergravity contributions.  This is one of
  many complications in understanding in detail the systematics of the
correspondence between the higher-loop supergravity diagrams and string diagrams.    Whereas the $\calR^4$ and related terms of the same dimension are
integrals over half the superspace,  terms with more derivatives are
 formally integrals
over a higher fraction of the superspace.   Each power of momentum is equivalent
 to two powers of $\theta$ so that terms with   less than eight powers of momentum acting on $\calR^4$ should be protected and may be determined in this
 manner.   This includes the $S^3\, \calR^4 \sim D^6\, \calR^4$ term which should therefore
  also be determined by similar considerations.
Whether it is possible to go beyond this and relate  terms in string
perturbation theory  at higher order in the momentum expansion   to
eleven-dimensional supergravity is much less obvious.

\subsec{Comparison of type IIA and IIB perturbation expansions.
}

An important constraint on the structure of the results of the
eleven-dimensional calculations that we will make use of  is a
relationship between the type IIA and IIB four-graviton scattering
amplitudes that holds up to and including two loops.

It is well known that the tree-level and one-loop four-graviton  amplitudes of
the type IIA and type IIB superstring theories are identical (ignoring the
parity-violating
part of the loop amplitude, which vanishes in topologically trivial backgrounds).
This property, which  is also true for  compactifications,
is not an obvious  consequence of the duality symmetries.  For
example, T-duality
(which  applies to all orders in   perturbation
theory as well as non-perturbatively) only
identifies  the two theories when one  is
 compactified on a circle and the other
on the inverse circle, whereas we are comparing
 the theories on circles of the same radius (which may, for example,
be infinite).
  The question is how far this generalizes to
 higher genus diagrams, which have not been explicitly evaluated?
Such  equality can be
 seen by considering the explicit construction of the
 four-graviton loops in the two theories.\foot{This subsection is based on conversations
 with Nathan
Berkovits.}

   Recall that the type
 IIB theory differs from that of the type IIA by a flip of sign in
 the GSO projection for the odd spin structure of the left-moving
 fermions while the
 right-moving fermions have identical GSO projections.  Therefore,
  loop amplitudes with external gravitons,
 or any other massless states in the \NSNS\ sector, differ only
 in the sign of the odd-odd spin structures --- those spin
 structures that are odd both in the left-moving and in the
 right-moving sectors (we will again ignore the odd-even spin structures which vanish in
 the topologically trivial backgrounds that we are considering).
Consider the scattering of gravitons with momenta
 $k^{(r)}_{\mu _r}$ ($r=1,2,3,4$) where
$\sum_{r=1}^4  k^{(r)}_\mu =0$,  and polarization vectors
 $h^{(r)}_{\mu _r\nu_r}$ which can be written in terms of left-moving
 and right-moving vectors a
 $h^{(r)}_{\mu _r\nu_r}=\sum_i h^{(r;i)}_{\mu _r} \tilde
h^{(r;i)}_{\nu_r}$.
For genus $l\ge 1$ these terms are associated with the
 product of two epsilon tensors, $\epsilon^{\mu _0\dots
 \mu _9}\,\tilde \epsilon^{\nu_0\dots
 \nu_9}$.  The tensor indices  can contract with the three
 independent external momenta, the left-moving and right-moving
 vectors in the polarization tensors or with each other.  At one loop
 there are no contractions between the two epsilon tensors
 so the odd-odd spin structures vanish and the two
 theories are identical.  At higher loops there are insertions of
 $2l-2$ supermoduli associated with picture changing.  Each one of
 these inserts a factor of $\partial X^\mu \bar\partial X^\nu$
 which leads to a total of  two  contractions of the form
 $\eta^{\mu \nu}$ between the two epsilon tensors. This is still not
 enough to allow the sixteen remaining indices of these
tensors to be saturated by
 the external momenta and polarizations.   When $l>2$
  there are
 more contractions between the epsilon tensors due to the higher
 number of picture changing operators, in which case the odd-odd
 spin structures give a non-zero contribution.  We conclude,
 therefore, that:
\smallskip
\noindent
  {\it The four-graviton amplitudes in the type IIA and IIB
 superstring theories are equal up to two
 loops, but not beyond.}
\smallskip
\noindent   A corollary is that
amplitudes with five  external gravitons are not
 equal at  two loops while those with more gravitons are not equal at one
 or two loops.
%%%%%%%%%%%%%
%%%%%%%%%%%%%%%%%%%%%%%%%%%%%%%%%%%%%%%%%%%%%%%%%%%%%%%%%%%%%%%%%%%%
\newsec{Momentum dependence of the one-loop
supergravity amplitude}

The one-loop amplitude describing the elastic scattering of two
gravitons in eleven-dimensional Minkowski space is given by
\refs{\rfGreenGutperleVanhove},
\eqn\eOneLoop{
A_4^{(1)} = {\kappa_{11}^4\over (2\pi)^{11}}\;\hK \
[ I(S,T)  + I(S,U) + I(U,T)], }
where  the function $I$
has the form of a Feynman integral for a box diagram in  massless scalar
$\varphi^3$ field theory,
\eqn\oneloopsd{I(S,T) = \int d^{11} q\,
{1\over q^2}{1\over (q+k_1)^2} {1\over (q+k_1+k_2)^2} {1\over (q-k_4)^2},}
and  $q_\mu $ ($\mu =0, \cdots,10$) is the eleven-dimensional loop momentum.  The
numerical coefficient in \eOneLoop\ follows the conventions of \refs{\rfBernDunbar}
(with a slight reshuffling of the powers of $(2\pi)^{11}$)
which will be convenient for later consideration of two-loop diagrams.

We want to consider this amplitude compactified  on
$M^9 \times \calT^2$ so that two components of the  loop momentum
are proportional to integer Kaluza--Klein charges
 ($l_1$ and $l_2$).  For simplicity, we will choose a kinematic
configuration in which the external gravitons have their polarizations and
momenta oriented in directions transverse to the two-torus.
After representing the propagators as integrals of Schwinger parameters
in the usual manner the compactified version of \oneloopsd\ can be written as
\eqn\capidef{ I(S,T)
= {1\over l_{11}^2\calV}\int \prod_{r=1}^4 d\sigma_r \int d^9q \
\sum_{\{l_1,l_2\}} e^{- G^{IJ}l_{I}l_{J}  \sigma -    \sum_{r=1}^4 p_r^2
\sigma_r},}
where $\sigma = \sum_{r=1}^4 \sigma_r$
 and   $p_r = q+\sum_{s=1}^r k_s$ are the momenta in the legs of the
loop. The Schwinger  parameters $\sigma_r$ have dimension $(length)^2$.

After a few manipulations
 \refs{\rfRussoTseytlin,\rfGreenCargese,\rfGreenGutperleVanhove}
each of the three terms in the  scalar integral \capidef\ can be rewritten as
\eqn\eIST{
I(S,T) = {2\pi^{9\over 2} \over l_{11}^2\calV} \int_0^\infty d\sigma \;
 \sigma^{-{3\over 2}}
\int_{\calT_{ST}} \prod_{r=1}^3 d\omega_r \ \sum_{\{ l_1, l_2\}} e^{- G^{IJ} l_{I}
l_{J}  \sigma -Q(S,T;\omega_r)\sigma }
}
where $Q(S,T;\omega_r)= -S \omega_1(\omega_3-\omega_2) - T (\omega_2-\omega_1)
(1-\omega_3)$.
The domain of integration indicated by
 $\calT_{ST}$  is defined by $0\leq \omega_1\leq \omega_2 \leq \omega_3 \leq 1$.
The other two terms in \capidef\
come from integration over the two remaining regions,
$\calT_{TU} : 0\leq
\omega_3\leq \omega_2 \leq \omega_1 \leq 1$ and  $\calT_{SU} : 0\leq
\omega_2\leq \omega_1 \leq \omega_3 \leq 1$.
  The integral \eIST\ is to be evaluated with $S, T<0$ where it
converges and then analytically continued to the physical region.  The
amplitude can be split into a momentum-dependent and a
momentum-independent part
\eqn\eISTSplit{
I(S,T)=I_o + I'(S,T)\ ,
}
where
\eqn\iodef{
I_o\equiv I(0,0)= {\pi^4\over l_{11}^2\calV} \int_0^\infty d\sigma
 \sigma^{-{3\over 2}} \sum_{\{ l_1,l_2\} } e^{-\pi G^{IJ} l_I l_J \sigma}, }
where $\sigma$ has been rescaled by a factor of $\pi$ in passing from
 \eIST\ to  \iodef.
This expression diverges for small $\sigma$, which is the cubic ultraviolet
 divergence of the scalar box diagram in eleven-dimensions.  We will
 regularize this divergence by introducing a cutoff on $\sigma$ so that
 $1/\Lambda ^2 \le \sigma$. It is convenient to  carry  out Poisson summations
 on $l_1$ and $l_2$ which replaces the Kaluza--Klein charges by winding
 numbers, $\hat l_1$ and $\hat l_2$.  The divergence is now isolated in  the
 zero winding number term, $(\hat l_1,\hat l_2)=(0,0)$.  The result is
\eqn\eizero{
I_o   ={\pi^3\over2l_{11}^3} \left( {4\pi\over 3}(\Lambda l_{11})^3  +
 \calV^{-{3\over2}} f_1(\Omega,\bar \Omega)\right), }
where $\Lambda^3$ is the regularized value of
the zero winding number term.   In balancing the dimensions in
this and other equations it is important to note that we have
defined all distances as dimensionless multiples of the Planck
distance, $l_{11}$.   In this convention  $\calV$ is
dimensionless  while the one-loop cutoff $\Lambda$ has dimension
 $(length)^{-1}$.
In addition to the one-loop contribution there is the freedom to
add the  local counterterm, $\delta^{(1)}S \sim l_{11}^{-3}c_1
 \int d^9x\sqrt{-G^{(9)}} \calV\, \calR^4$,
which adds a term,
\eqn\oneadd{\delta A_4^{(1)}={ \kappa_{11}^4\over (2\pi)^{11}}\,
\hat K\, \delta
 I_o,}
 to the  amplitude, where
\eqn\countsed{\delta I_o ={\pi^3\over 2l_{11}^{3}} c_1 ,}
and $c_1$ is an arbitrary coefficient that will shortly
be given a $\Lambda$-dependent value.

The $\Lambda$-independent term in \eizero\ has
coefficient $f_1(\Omega,\bar\Omega)=2\zeta(3) E_{3/2}$,
 where the
Eisenstein series $E_s$ is defined by
\eqn\eisendef{2\zeta(2s) E_s = \sum_{(m,n)\ne (0,0)} {\Omega_2^s
\over |m+ n\Omega|^{2s}}.}
The  volume-dependence, $\calV^{-3/2}$,  of this term  means that it
vanishes in the eleven-dimensional limit, $\calV\to \infty$.
However, this is the only term that survives in the limit that corresponds to the
decompactified type IIB theory,
$r_B\to \infty$ with fixed $e^{\phi^B}$,
while the cutoff dependent term in \eizero\ gives vanishing
contribution.
 The complex structure of $\calT^2$ is to be identified  with the
  complex IIB scalar field,
 $\Omega = \Omega_1+ i\Omega_2$.  Expanding $E_{3/2}$ for large
 $\Omega_2$ (small type IIB coupling, $e^{\phi^B}$) gives
 \eqn\eexpan{2\zeta(3) E_{3\over 2} = 2\zeta(3) e^{-3\phi^B/2} +
 {2\pi^2\over 3} e^{\phi^B/2} + {\rm non-perturbative}.}
 The successive  terms in this expansion can be identified with
  tree-level, one-loop, and non-perturbative terms in the coefficient of
  the type IIB string theory
$\calR^4$  interaction.  The non-perturbative terms have the
  form of an infinite series of D-instanton terms where each
  charge-$K$ D-instanton contribution has an infinite series of
  perturbative fluctuations.

The total one-loop contribution to the amplitude comes from the
combination $I_o+\delta I_o$ (the sum of \countsed\ and  \eizero).
The dominant term in the large-$\calV$ limit is proportional to
$(4\pi(\Lambda l_{11})^3/3  +c_1)$ and is independent of $\calV$ so in
the string theory parameterization this term
is independent of the dilaton and arises
 from  one  string loop in the type IIA
theory.     Although this coefficient  is not determined  by
the physics of  quantized eleven-dimensional
supergravity, it is determined by insisting that the four-graviton
interactions in the type
IIA and  type IIB  effective string actions should be equal when
the radii $r_A$ and $r_B$ are equal.  As argued in section 2.3
this is known property of string perturbation theory up to and
including two loops.  More explicitly, the nine-dimensional effective
actions that give rise to the momentum independent part of
$I+\delta I_o$ have $\calR^4$ terms that are expressed as
\eqn\twoactm{S_{\calR^4} =
{1\over 3\cdot (4\pi)^7\;l_{11}}\int d^9x\, \sqrt{-G}\; \C|V\,  \calR^4
\left(2\zeta(3)  \C|V^{-{3\over 2}}E_{3\over 2}(\Omega,\bar\Omega) + c_1 +{4\pi\over 3} (\Lambda l_{11})^3 \right),}
which can be written in string theory coordinates and expanded for
small string coupling constant in the form
\eqn\twoacti{
\eqalign{
S_{\calR^4}  &={1\over 3\cdot (4\pi)^7\;l_s}\int d^9x\,\sqrt{-g^B}\; r_B\,
\calR^4 \left(2\zeta(3) e^{-2\phi^B}+{2\pi^2\over 3} + {c_1 +4\pi (\Lambda
    l_{11})^3/3\over r_B^2}+\cdots  \right)\cr
&={1\over 3\cdot (4\pi)^7\;l_s}\int d^9x\, \sqrt{-g^B}\;r_A\, \calR^4
\left(2\zeta(3) e^{-2\phi^A}+{2\pi^2\over 3}{1\over r_A^2} +
c_1+{4\pi\over3}  (\Lambda l_{11})^3 +\cdots  \right),
}}
in the type IIA and type IIB theories, respectively
(ignoring the non-perturbative contributions).
It follows that the only consistent value for  $c_1$ which equates the IIA and IIB
expressions is
\eqn\regcon{c_1=  {2\pi^2\over 3}  -  {4\pi\over3} (\Lambda l_{11})^3.}
This is the value which is also consistent with supersymmetry
\refs{\rfGreenVanhoveMtheory}.

The momentum dependence of $I(S,T)$ in~\eIST\ is contained in the finite term
 $I'(S,T)$ in~\eISTSplit. We will separate the term with zero Kaluza--Klein
 momenta,  $I^0(S,T)$ $(l_1=l_2=0)$,  from the rest by writing
\eqn\iprimedef{
{I}'(S,T)= I^0(S,T)+ \sum_{n=2}^\infty I_n(S,T)\ .}
The  term $I^0(S,T)$, which contains the non-analytic contribution to the
 amplitude \refs{\rfRussoTseytlin,\rfGreenCargese} has the form in $d$ dimensions,
\eqn\ddim{
\eqalign{
l_{11}^2\calV\, I^{0}_d(S,T)  &= 2\pi^{d\over 2} \int_0^\infty d\sigma \;
\sigma^{3-{d\over2}} \int_{\calT_{ST}} \prod_{r=1}^3 d\omega_r \; \left(e^{-Q(S,T;\omega_r) \sigma}-1\right)\cr
&=2\pi^{d\over2}\Gamma(4-{\textstyle{d\over2}}) \int_{\calT_{ST}}\prod_{r=1}^3   d\omega_r
Q(S,T;\omega_r)^{d-8\over 2} \cr
&= 2\pi^{d\over 2} \Gamma(4-{\textstyle{d\over2}})
\left(-\GST\right)^{d-8\over 2},
}}
where  $\GST^n$ is defined by
\eqn\calsdef{
\GST^n = \int_{\calT_{ST}}\prod_{r=1}^3  d\omega_r (-Q(S,T;\omega_r))^n.}
Similarly, $\GTU^n$ and $\GUS^n$ will be defined by cyclically permuting $S$,
$T$ and $U$ in the function $Q$.
Specializing to $d=9$ gives,
\eqn\eNonAnalytic{
l_{11}^2\calV\, I^0(S,T) \equiv - 8 \pi^5 (-\GST)^{\half}
=-8 \pi^5 \int_{\calT_{ST}} \prod_{r=1}^3 d\omega_r (Q(S,T;\omega_r))^\half.}

The terms in \iprimedef\ with non-zero Kaluza--Klein charge, $I_n$,
 are homogeneous polynomials in $S$ and $T$ of degree $n$,
\eqn\nonzek{\eqalign{
l_{11}^2\calV\, I_n(S,T) &=2 \pi^{9\over 2}\, {\GST^n \over n!}\,  \int_0^\infty
 {d\sigma \over \sigma^{{3\over 2} -n} }
 \sum_{( l_1, l_2) \ne (0,0)} e^{- G^{IJ} l_{I} l_{J}
  \sigma } \cr
& = 4 \pi^{9\over 2} \Gamma(n-\half)  \zeta(2n-1) l_{11}^{2n-1}\calV^{n-\half}\,
{\GST^n\over n!}\,  E_{n-\half} (\Omega,\bar \Omega) \ .
}}
The Eisenstein series' that enter this expression have the
large-$\Omega_2$ expansion,
\eqn\esexp{  E_{n-\half}(\Omega,\bar\Omega) =
e^{-(2n-1)\phi^B/2} + {\sqrt\pi \Gamma(n-1)\zeta(2n-2)\over
 \Gamma(n-\half)\zeta(2n-1)} e^{(3-2n)\phi^B/2} + {\rm non-pert.}}
The first term will  contribute to the tree-level amplitude when
expressed in string coordinates and the second term is a $n$-loop
contribution.  All the other terms are non-perturbative D-instanton effects.

Using the expansion \esexp\  and putting all the terms together, the  complete
expression for the $\calR^4$ term in the one-loop effective action for
eleven-dimensional supergravity in nine-dimensions is given by~\eMCorrections\
with the function $h$ defined by the amplitude $A_4$ in~\eOneLoop\ where
\eqn\eMall{\eqalign{
 I(S,T)+\,& I(T,U)\, + I(U,S) =I_o - {4 \pi^5\over l_{11}^2\calV}
(-\calW)^{\half} \cr
&  + 4\pi^{9\over 2} \sum_{n=2}^\infty l_{11}^{2n-3}    {{\calW}^n \calV^{n-{3\over
2}} \over n!}   \left[ \Gamma(n-\half)\zeta(2n-1) \left({R_{10} \over R_{11}}
\right)^{n-\half} \right. \cr
&\left. +  \sqrt \pi \Gamma(n-1) \zeta(2n-2) \left({R_{10} \over R_{11}}
 \right)^{{3\over  2} - n}\right]\cr & + {\rm non-perturbative}\ , \cr}}
and
\eqn\calwdef{
{\calW}^n = \GST^n+\GTU^n+\GUS^n.}
There is no $n=1$ term  after adding the contributions of $I(S,U) $ and
$I(T,U)$ to $I(S,T)$
since the   linear symmetric combination vanishes after using the mass shell
condition.

The non-analytic term~\eNonAnalytic\ in nine-dimensional M-theory comes from
the same massless thresholds that arise in either nine-dimensional type II
string theory \refs{\rfGreenCargese,\rfRussoTseytlin} and have square root
branch cuts of the form $(-S)^{1/2}$.

The two infinite series of terms on the right-hand side of \eMall\
have very obvious origins from the dimensional reduction of the
massless one-loop normal threshold of the eleven-dimensional loop.
The first infinite series is a series of ascending powers of $R_{10}$.  Although
such terms appear to give singular behaviour in the ten-dimensional IIA
decompactification limit,  $R_{10}  \to \infty$, the series actually
sums up to give the correct threshold behaviour of the
ten-dimensional theory, schematically of the form $S\ln (-S)$.  More
explicitly, the sum gives
\eqn\sumlogs{\eqalign{
- {4 \pi^5\over l_{11}^2\calV}& (-\calW)^{\half} + 4\pi^{9\over2} \sum_{n=2}^\infty
l_{11}^{2n-3}
{{\calW}^n \calV^{n-{3\over 2}} \over  n!} \Gamma(n-\half)\zeta(2n-1)
  \left({R_{10} \over R_{11}} \right)^{n-\half} \cr
& = - {4 \pi^5\over l_{11}^3\calV}
  \sum_{r\in \ZZ} \left({r^2\over R_{10}^2} - l_{11}^2\C|W \right)^\half -
{1\over l_{11}^3R_{11}R_{10}^2}4\pi^5 \zeta(-1) \Gamma(-\half).
}}
The last term on the right-hand side cancels against the second term  in the
expansion of $E_{3/2}(\Omega,\bar\Omega)$ in $I_o$  \eizero\ (using \eexpan).
 The net result is that the sum of the first infinite series of terms in \eMall\
is given by
\eqn\massth{
\int_{T_{st}} \prod_{r=1}^3 d\omega_r \left[- {8 \pi^2\over l_{11}^3\calV}
\sum_{r\in \ZZ} \left({r^2\over R_{10}^2} +l_{11}^2Q(S,T;\omega_r)\right)^\half\right].}
The sum  over $r$  can be evaluated in the large-$R_{10}$ limit by approximating it
 by an integral by letting  $r/R_{10}\to y$,
\eqn\intyy{\eqalign{
\lim_{R_{10}\to\infty} - {4 \pi^5\over l_{11}^3\calV}
  \sum_{r\in \ZZ} \left({r^2\over R_{10}^2}- l_{11}^2\C|W
  \right)^\half&=
    -{8\pi^5\over l_{11}^3R_{11}} \int_0^\infty dy (y^2-l_{11}^2\C|W)^\half\cr
& = {2\pi^5\over l_{11}R_{11}} \C|W \left(\ln(-\C|W) -2 \right) ,}}
where  a constant has been absorbed into the implicit scale of the
logarithm.
This cancels  out when the $(S,T)$, $(T,U)$ and $(U,S)$
contributions are added.  The result is that the
 series sums up to the expected massless logarithmic threshold in ten
 dimensions of the form
\eqn\eLogCut{\eqalign{ &
 \calR^4(
\GST\ln \GST +  \GTU\ln \GTU + \GUS \ln \GUS)
  \cr
& \sim   \calR^4( \calgst\ln \calgst +
\calgtu\ln \calgtu + \calgus \ln \calgus)\ .\cr}}
where the expressions $\calgst,\calgtu$ and $\calgus$ are defined in terms of the
Mandelstam invariants of string theory,
using the relation between the
M-theory and string theory  Mandelstam invariants, \relman.
 Both~\eNonAnalytic\ and~\eLogCut\ have
imaginary parts corresponding to the massless normal thresholds determined by
unitarity.  However, the real parts, which might have given rise to arbitrary
constants,  are here fixed to precise values.

In addition to the threshold term \eLogCut, the tree-level part of
$I_o$  also survives the limit $R_{10}\to \infty$ in the  type IIA
amplitude, but the D-instanton terms are infinitely suppressed and
disappear.

The second series of terms on the right-hand side of~\eMall\ is an
ascending series of powers of $R_{11}$.  In the ten-dimensional
decompactification limit, $R_{10}\to \infty$, this sums to a series
of  massive logarithmic thresholds of the ten-dimensional theory,
\eqn\secser{\eqalign{
 2 \pi^5\sum_{n=1}^\infty  l_{11}^{2n-3} {\calW^n \calV^{n-{3\over 2}} \over
n!} &   \Gamma(n-1) 2\zeta(2n-2) \left({R_{10} \over R_{11}} \right)^{{3\over
2} - n} \cr
=& {2\pi^5\over l_{11}^3R_{11}} \sum_{r\neq 0} \left({r^2\over R_{11}^2}-
l_{11}^2\C|W \right) \left(\ln\left( 1-{R_{11}^2\over r^2}
l_{11}^2\C|W \right)-2 \right)\ .\cr}}
This is a series of thresholds for the massive Kaluza--Klein states of
 M-theory on a circle of radius $R_{11}$ which is to be added to the
  massless threshold that   comes from ~\intyy.
This sum can be evaluated in the decompactification limit $R_{11}\to
\infty$ by approximating $r/R_{11}$ by a continuous
variable. Including  the massless threshold term
\eLogCut, the sum reduces to the eleven-dimensional threshold,
\eqn\elevcut{
\eqalign{
l_{11}^2\calV\, I=&2\pi^{9\over 2} \int_0^\infty d\sigma
\sigma^{3-{11\over 2}}\int \prod_{r=1}^3 d\omega_r e^{-\sigma
  Q(\omega_r)}  \cr
  =& {8\pi^5\over 3} \left[
  (-\GST)^{3\over 2}+(-\GTU)^{3\over 2}+(-\GUS)^{3\over 2}\right].\cr
}}
The tree-level type IIA term vanishes in the $R_{11}\to\infty$ limit.

Rewriting the result of the one-loop calculation \eMall\
 in terms of the string theory
parameters gives  terms in the effective action of
nine-dimensional  IIA and IIB string theory with derivatives acting on
 $\calR^4$ that can be written in terms of the one-loop amplitude
\eOneLoop\  as\foot{These expressions correct certain coefficients in
 the corresponding formulas of \refs{\rfGreenCargese}.},
\eqn\estringa{
\eqalign{   A_4^{(1)}=& (4 \pi^8 l_{11}^{15}r_A^{-1}) \;\hK\,
r_A\left[
 2\zeta(3)e^{-2\phi^A}+ {2\pi^2\over 3 r_A^2} + {2\pi^2\over 3}
 -8\pi^2 r_A\; l_s (-\calw)^{\half}\right.\cr
 &\left. +8\pi^{3\over 2}\sum_{n=2}^\infty \left(\Gamma(n-\half)\zeta(2n-1) 
  {r_A^{2(n-1)}\over n!}(l_s^2 {\calw})^n 
\right.\right. \cr & \left.\left. + \sqrt \pi \Gamma(n-1)
 \zeta(2n-2) {e^{2(n-1)\phi^A}\over n!}(l_s^2{\calw})^n \right)
 \right]   + {\rm non-perturbative}, \cr}}
     or
\eqn\estringb{
\eqalign{
 A_4^{(1)} = &
   (4 \pi^8 l_{11}^{15}r_B)\;
  \hK\,  r_B \left[ 2\zeta(3) e^{-2\phi^B} + {2\pi^2 \over 3} +
 {2\pi^2 \over 3}  {1\over r_B^2} - {8\pi^2
 \over r_B^3} l_s (-\calw)^{\half}\right.\cr
&\left.   + 8\pi^{3\over2}\sum_{n=2}^\infty \left(\Gamma(n-\half)\zeta(2n-1){1
 \over n!} {(l_s^2{\calw})^n \over r_B^{2n+2}} \right.\right. \cr
&  \left.\left. + \sqrt \pi \Gamma(n-1) \zeta(2n-2)
{e^{2(n-1)\phi^B}\over n!} {(l_s^2{\calw})^n
  \over r_B^{2n+2}}\right)\right]   +{\rm non-perturbative}\ ,
}}
where
\eqn\smalls{
{\calw}^n = ({\calgst})^n + ({\calgtu})^n + ({\calgus})^n.}
The overall factor of  $4\pi^8l_{11}^{15}\, r_A^{-1} = 4\pi^8l_{11}^{15}\, r_B$
that has been factored out in these expressions cancels with a factor in
 the measure in transforming the effective action from
 eleven-dimensional supergravity coordinates to  string coordinates. This
 makes it easy to see the dependence of the effective interactions on the
 string-frame radius and the dilaton in
 in the remaining factors in \estringa\ and \estringb.

The infinite series' of  terms in the IIA theory are related by T-duality to
the series' in the IIB theory.  However, these terms appear asymmetrically
between the two theories in  \estringa\ and \estringb. In particular, all the
terms in the series' in the IIB action vanish in the limit $r_B\to \infty$
which is not true for the IIA   series' in the IIA decompactification
limit. Since   we saw in section 2.3 that the four-graviton amplitudes in the
IIA and IIB string theories are identical up to and including the
contributions from two string loops there must be some more contributions
that correct for this asymmetry. We will be concerned particularly with the
$n=2$ terms on the right-hand side of the IIA action in \estringa,
\eqn\twonterm{\eqalign{ A_4^{(1)\,n=2}
=&4 \pi^8 l_{11}^{19}
 e^{-{4\over 3}\phi^A} \times \pi^2\left(2\zeta(3)  r_A^2 + 4\zeta(2) e^{2
  \phi^A}\right)\ (\calw)^2 \hK\cr
=& {4 \pi^{10}\over  6!}l_{11}^{19}e^{-{4\over 3}\phi^A} \left(2\zeta(3)
  r_A^2  + {2\pi^2\over 3} e^{2 \phi^A}\right)\ (s^2+t^2+u^2) \hK ,\cr}}
where  we have used
\eqn\wsqare{(\calgst)^2 = \int_{\calT_{st}} \prod_{r=1}^3 d\omega_r
  \left(-s \omega_1(\omega_3-\omega_2) - t (\omega_2-\omega_1)
    (1-\omega_3)\right)^2  = {1\over 7!} ( 4s^2 + 4t^2 +  2st).}
The expression \twonterm\ has a dependence on the dilaton that is
characteristic of contributions in IIA string theory at  one and two loops.
Since the type IIA and type IIB string perturbation
 theories are identical up to two loops  these IIA terms  must
 be matched by identical terms in the IIB theory
(with $r_A\to r_B$ and $\phi^A\to \phi^B$).   We will see in the next
section that
these missing contributions to the IIB action arise from the
compactification of two-loop terms in  eleven-dimensional
supergravity.

%%%%%%%%%%%%%%%%%%%%%%%%%%%%%%%%%%%%%%%%%%%%%%%%%%%%%%%%%%%%%%%%%%%%%%%%%%%%%

\newsec{The two-loop supergravity amplitude}

The evaluation of two-loop amplitudes in eleven-dimensional quantum supergravity would
 normally be a
formidable task.  However,
it is known from the work of \rfBernDunbar\ that
the two-loop four-graviton amplitude in
 maximally supersymmetric supergravity continues to have the feature
that it can be written in terms of scalar field theory diagrams.  The
fact that the two-loop amplitude has such a simple expression was
motivated in \rfBernDunbar\ in dimensions $\le 10$ by use of the
Kawai--Llewelyn--Tye (KLT) rules for
constructing closed-string amplitudes out of open-string amplitudes
\refs{\rfKawaiLewellenTye}.  This was shown to imply that the two-loop
amplitude in the low energy
supergravity theory in $d$ dimensions with maximum  supersymmetry is
given in terms of the two-loop amplitude of
supersymmetric Yang--Mills theory with maximal
supersymmetry.
These rules were then independently
derived by using unitarity in all channels.   In eleven
dimensions supergravity  is  not the low energy limit of a string theory so the
strategy for determining the two-loop amplitude
has to be a little different.  In that case the expression may be
determined by the requirement of unitarity and can  also be  checked by
 the requirement that it reduce to the lower-dimensional
expressions upon trivial dimensional reduction.

%%%%%%%%%%%%%%%%%%%%%%%%%%%%%
\ifig\fthree{The `$S$-channel'
scalar field theory diagrams that contribute to the
two-loop four-graviton amplitude of eleven-dimensional supergravity.
(a)  The $(S,T)$ planar diagram, $I^P(S,T)$; (b) The $(S,T)$
non-planar diagram, $I^{NP}(S,T)$. }
{\epsfbox{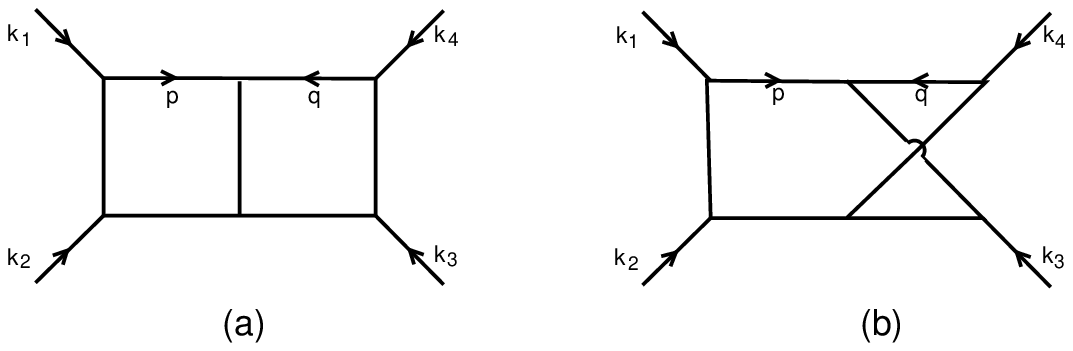}}
%%%%

 The result is that  the two-loop four-graviton amplitude,
$A^{(2)}_4(S,T,U)$,   is given in
terms of the sum of particular diagrams of $\varphi^3$
scalar field theory illustrated in \fthree.  These are the planar
diagram, $I^P(S,T)$, and the non-planar diagram, $I^{NP}(S,T)$, together
with the other diagrams obtained by permuting the external particles.
The complete expression for the amplitude is (with same conventions as in
\refs{\rfBernDunbar})
\eqn\eberndun{
A_4^{(2)}=\! i{\kappa_{11}^6\over (2\pi)^{22}}\; \hK\!
\left[S^2 \left(I^P(S,T)+\!  I^P(S,U)
+\! I^{NP}(S,T) +\! I^{NP}(S,U)\right)+ perms. \right],}
where $perms$  signifies the sum of terms with permutations of $S,T$
and $U$.  This expression has an overall factor of $\calR^4$ together with
four powers of the momentum multiplying the loop integrals which means
that these diagrams are much less divergent than they would
naively appear.  The loop integrals are given by
\eqn\eIP{
I^P(S,T)= \int d^{11}pd^{11}q {1\over p^2 (p-k_1)^2 (p-k_1-k_2)^2 (p+q)^2 q^2
  (q-k_3-k_4)^2 (q-k_4)^2}
}
and
\eqn\eINP{
I^{NP}(S,T)= \int d^{11}pd^{11}q {1\over p^2 (p-k_1)^2  (p+q)^2(p-k_1-k_2)^2 q^2
  (p+q+k_3)^2 (q-k_4)^2}\
}
which have ultraviolet divergences of order $(momentum)^8$ that will need to be
regularized.

In addition to these two-loop diagrams there is a contribution to the amplitude
from the one-loop diagram of \ftwo(b), which is a triangle diagram
 in which there is one insertion of the  linearized one-loop
counterterm.  Together with two-loop counterterm of \ftwo(c),  this
 will give an additional contribution, $\delta A_4^{(2)}$, to the amplitude.

%%%%%%%%%%%%%%%%%%%%%%%%%%%%%%%%%%%%%%%%%%%%%%%%%%%%%%%%%%%%%%%%%%%%%%%%%%

\subsec{Evaluation of the two-loop amplitude on $\calT^n$}

We shall now consider the leading contribution to the derivative
expansion arising from these two-loop diagrams when compactified
on $\calT^2$.  As discussed earlier, this will
contribute to the  $D^4 \calR^4$ interaction. For convenience
 our considerations will be restricted to situations in which the
polarization tensors and momenta of the gravitons are in
directions transverse to torus and covariantize the final result.
We will first be slightly more general and consider the
case of compactification on an
$n$-torus $\calT^n$ with metric $G_{IJ}$ and
volume $\calV_n$, in which case the  planar diagram with external momenta
$k_r$ $r=1,\dots,4$ is given by the expression,
\eqn\eintplan{\eqalign{
I^P(S,T) = &{1 \over l_{11}^{2n}\calV_n^2} \sum_{\{m_I,n_I\}} \int d^{11-n}p\;
d^{11-n}q\cr
& \int \prod_{r=1}^7 d\sigma_r \ e^{-\left[G^{IJ} \left(\sigma m_Im_J + \lambda
n_I n_J + \rho (m+n)_I (m+n)_J \right) + \sum_{r=1}^7 K_r \sigma_r \right]},
}}
where $I,J=1,2$ label the directions in $\calT^n$.  The vector $K_r$ is defined by
\eqn\kldef{K_r = (p,p-k_1,p-k_1-k_2,q,q-k_4,q-k_3-k_4,p+q),}
and
\eqn\schwindef{
\sigma=\sigma_1+\sigma_2+\sigma_3,\qquad \lambda=\sigma_4+\sigma_5+\sigma_6,
\qquad\rho=\sigma_7.}
The  non-planar diagram  is given by
\eqn\eIntegralNonPlanar{\eqalign{
I^{NP}(S,T) = & {1 \over l_{11}^{2n}\calV_n^2} \sum_{\{m_I,n_I\}}\int d^{11-n}p\;
d^{11-n}q\cr
&\quad  \int \prod_{r=1}^7 d\sigma_r \ e^{-\left[G^{IJ} \left(\sigma m_Im_J +
      \lambda n_I n_J + \rho (m+n)_I (m+n)_J \right) + \sum_{r=1}^7
  K^{\prime 2}_r \sigma_r \right]},\cr}
}
where
\eqn\nonplan{K'_r=(q, q-k_4, p, p-k_1, p-k_1-k_2, p+q, p+q+k_3),}
and
\eqn\schr{\sigma=\sigma_1+\sigma_2, \qquad\lambda=\sigma_3+\sigma_4+\sigma_5,
\qquad \rho=\sigma_6+\sigma_7.}

The loop momentum integrals are performed in the standard manner by
completing the squares in the exponent followed by gaussian
integration.  We are envisioning introducing some sort of  cutoff at
large momenta by imposing a lower limit to the  range of integration of
 the Schwinger parameters. The precise details will be clarified following
suitable changes of variables below. Ignoring these for now,
the resultant expressions for the planar
and non-planar loops are,
  \eqn\enonpll{\eqalign{
 &I^P(S,T) ={\pi^{11-n}\over l_{11}^{2n}\calV_n^2} \sum_{\{m_I,n_I\}}
 \int_0^\infty d\sigma d\lambda d\rho {\sigma^2\lambda^2\over
 \Delta^{11-n\over 2}} e^{-G^{IJ} \left(\sigma m_Im_J + \lambda n_I n_J + \rho
 (m+n)_I (m+n)_J \right)} \cr
& \int^1_0dv_2dw_2\int^{v_2}_0dv_1 \int^{w_2}_0dw_1
 e^{T{\sigma\lambda\rho\over
 \Delta}(v_2-v_1)(w_2-w_1)+S[{\sigma\lambda\rho\over \Delta}
 (v_1-w_1)(v_2-w_2)+\sigma v_1(1-v_2)+\lambda w_1(1-w_2)]},}}
and
\eqn\enonnon{\eqalign{
&I^{NP}(S,T) ={\pi^{11-n}\over l_{11}^{2n}\calV_n^2}
\sum_{\{m_I,n_I\}}\int_0^\infty d\sigma d\lambda d\rho {2\sigma\lambda^2\rho
\over \Delta^{11-n\over 2}} e^{-G^{IJ} \left(\sigma m_Im_J + \lambda n_I n_J +
\rho (m+n)_I (m+n)_J \right)}\cr
& \int^1_0du_1dv_1dw_2\int^{w_2}_0dw_1 e^{T{\sigma\lambda\rho\over
 \Delta}(w_2-w_1)(u_1-v_1) +S[{(\sigma+\rho)\lambda^2\over \Delta} w_1(1-w_2)
 +{\sigma\lambda\rho\over\Delta}( w_1(1-u_1) +v_1(u_1-w_2))]}}}
(where the variables $u_1$, $v_1$, $v_2$, $w_1$ and $w_2$ are rescalings of
$\sigma_i$).   These expressions can
be expanded in powers of $S,T$ and $U$ in order to determine their
 contributions to higher derivatives acting on $S^2\, \calR^4$.

The leading term in the low energy expansion (of order $S^2\, \calR^4$)
is obtained by
setting the external momenta to zero so
that $S$, $T$ and $U$ are set equal to zero in $I^P$ and
$I^{NP}$.
After summing these two zero-momentum contributions followed by a sum  over all the
diagrams required by Bose symmetrization the result is
\eqn\esumdi{
I^{P}(0)+I^{NP}(0)={\pi^{11-n} \over 3\;l_{11}^{2n}\calV_n^2}
\sum_{\{m_I,n_I\}}\int_0^\infty
d\sigma d\lambda d\rho\; {1\over \Delta^{7-n\over2}}
\; e^{- G^{IJ} \left(\sigma m_Im_J + \lambda n_I n_J + \rho (m+n)_I (m+n)_J
\right)},
}
which is symmetric in   the parameters
$\sigma,\lambda$ and $\rho$.  The integration  in \esumdi\
is  divergent for every value of $m^I,n^I$ when $\Delta \sim 0$, which requires
 at least  two of the parameters $\lambda,\rho,\sigma$ to approach
 zero simultaneously. The sums contribute additional  divergences, which makes this
 representation of the amplitude rather awkward to analyze.

As in the case of the one-loop amplitude it is convenient to
analyze the divergences after
performing a Poisson resummation over the Kaluza--Klein modes, $m_I,n_I$,  which
transforms them into winding numbers, $\hm_I, \hn_I$, and also to
redefine the Schwinger parameters by,
\eqn\defsn{\hsigma = {\sigma\over \Delta},\qquad \hlambda= {\lambda\over
\Delta}, \qquad \hrho = {\rho\over \Delta},}
where
\eqn\deldef{\Delta = \sigma\lambda+\sigma\rho+\lambda\rho =
{\hdelta}^{-1}
=(\hsigma\hlambda+\hsigma\hrho+\hlambda\hrho)^{-1}.}
The amplitude \esumdi\ becomes
\eqn\esum{
I^{P+NP}(0) ={\pi^7\over 3}
 \sum_{\{m_I,n_I\}} \int_0^\infty d\hsigma\, d\hlambda\, d\hrho\,
\hdelta^{1/2}\, e^{-\pi E_w} \ ,
}
where  the exponent is defined by
\eqn\eEw{
E_w(\hsigma,\hlambda,\hrho) = G_{IJ} \left(\hlambda \hm_I \hm_J +
\hsigma \hn_I \hn_J
+ \hrho (\hm+\hn)_I (\hm + \hn)_J
\right),
}
and is a function of the winding numbers.  The parameters $\hsigma$, $\hlambda$ and
$\hrho$ will be referred  to as `winding parameters'.
 The  classification of the
divergences is simplified in the winding number basis.
  For example, the sector in which all the
winding numbers vanish diverges at the end-point where
 all of the winding parameters reach their upper limits.
 This term  is independent of the metric $G_{IJ}$  and is the
primitive two-loop
 divergence.  There are many sectors that contribute to subleading
divergences. The simplest  examples are those  sectors
in which the winding numbers
conjugate to a particular winding  parameter  vanish.  In those cases
the integral  diverges
at the endpoint where that parameter reaches its upper limit,
 which gives a sub-leading divergence.  For example,
the $\hsigma$ integral diverges
 in the $\hat n_I=0$ sector and behaves as $\Lambda^3$  if
 $\hsigma$ is cut off at the value $\Lambda^2$ (that
 was introduced in order to cut off the one-loop winding parameter).
Sectors with less than $n$ vanishing winding numbers give
  non-divergent contributions which are independent of any cutoff.

%%%%%%%%%%%%%%%%%%%%%%%%%%%%%
\ifig\ffour{ The domain of integration over the parameters $\tau_1$
and $\tau_2$,  bounded by the thick line, is the fundamental domain
of $\Gamma_0(2)$.}
{\epsfbox{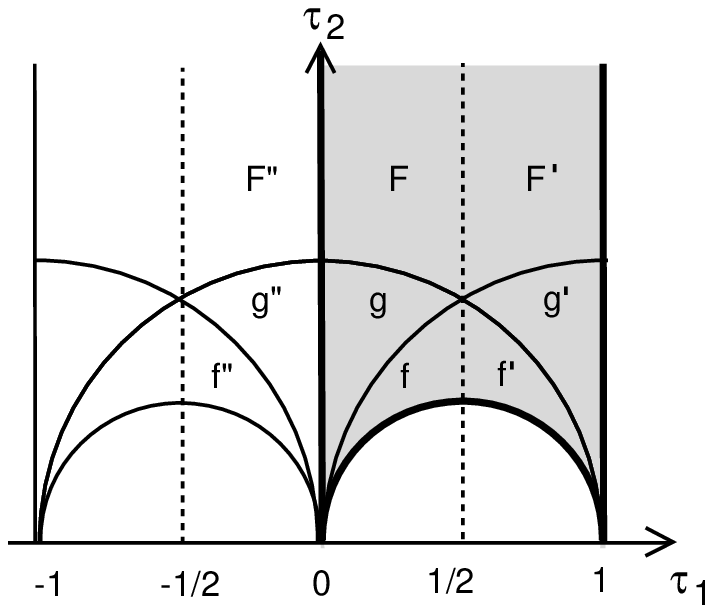}}
%%%%
A more complete analysis of the
divergences  is greatly  facilitated  by the observation that
the integrand possesses a secret $SL(2,\ZZ)$  symmetry that is not at
all apparent in the $\hlambda,\hrho,\hsigma$ parameterization.
This  symmetry is made  manifest by redefining  the integration variables in \esum\
 so that the parameters, $\hrho, \hlambda$ and $\hsigma$,
are replaced by the dimensionless volume, $V$, and complex
structure, $\tau = \tau_1 + i\tau_2$, of a two-torus, $\hat \calT^2$, defined
by
\eqn\modef{
\tau_1 = {\hrho\over \hrho+\hlambda}, \qquad \tau_2={\sqrt{\hat\Delta} \over
  \hrho+\hlambda},\qquad V=l_{11}^2\sqrt{\hDelta}\ .}
    The jacobian for the change of variables   from $(\hsigma,\hlambda,\hrho)$
    to  $(V,\tau)$ is
\eqn\jacob{d\hlambda d\hsigma d\hrho = 2l_{11}^{-6} \,dV\, V^2\, {d^2\tau \over \tau_2^2},}
where $d^2\tau=d\tau_1 d\tau_2$. It is easy to see how the domain of
integration of the Schwinger variables translates into the integration domain
for $V$ and $\tau$.  The volume $V$ is integrated over $[0,\infty]$ and the
domain of integration of $\tau$ is the fundamental domain of the $\Gamma_0(2)$
sub-group of $SL(2,\ZZ)$ (the shaded region in~\ffour),
\eqn\domsdef{
\calF_{\Gamma_0(2)} = \left\{0\leq \tau_1\leq 1, \tau_2^2+\left(\tau_1 -
    {1\over2} \right)^2\geq {1\over 4} \right\},}
which consists of the sectors $F\oplus F'\oplus g\oplus g' \oplus f \oplus
f'$.  As is clear from the \ffour\ this domain covers precisely three copies
of  $\calF = F\oplus F^{\prime\prime}$, the fundamental domain of
$SL(2,\ZZ)$. More concretely, in terms of the conventional generators of
$SL(2,\ZZ)$:\foot{Which are the translation  $T$ :  $\tau \to \tau+1$ and  the
  inversion $S$:  $\tau \to -1/\tau$.}  region $g$ is mapped into
$F^{\prime\prime}$ by $S$;  region $g'$ is mapped into $F$ by $ST^{-1}$;
region $f$ is mapped into $F$ by $TS$; region $f'$ is mapped into
$F^{\prime\prime}$ by $T^{-1}ST^{-1}$; region $F'$ is mapped into
$F^{\prime\prime}$ by $T^{-1}$.

Substituting the  change of variables \modef\ into  the integral \esum\
gives
\eqn\eintom{
I^{P+NP}(0) = {2 \pi^7\over l_{11}^8} \sum_{\{\hm_I,\hn_I\}} \int_0^\infty dV
V^3 \int_\calF  {d^2\tau\over \tau_2^2} e^{-\pi{V G_{IJ}\over
    l_{11}^2\tau_2} \left[(\hm + \tau \hn)^I(\hm + \bar \tau \hn)^J \right]} \
.
}
The integrand is precisely that which arises in
one-loop diagrams in string theory compactified on $\calT^n$ where
the Lorentzian lattice is usually defined by
\eqn\eLattice{
(\tau_2)^{n/2} \Gamma_{(n,n)}(G_{IJ}) = R_{11}^n \sum_{(\hat m_I,\hat n_I)\in
  \ZZ^2} e^{-\pi{V G_{IJ}\over l_{11}^2\tau_2} \left[(\hm + \tau \hn)^I(\hm +
    \bar \tau \hn)^J \right]}.
}
In writing the integral \eintom\ we have used the fact that  the
 integrand  is invariant under $SL(2,\ZZ)$ transformations to equate it
to three times  the integration over a single copy of $\calF$.
This invariance of the integrand  can be seen by checking its transformations
 under  $T$ and $S$  (with $V$ being inert) which have
a simple interpretation in terms of the original winding parameters.
 The  $T$ transformation is given by
\eqn\eTTrans{
 \hlambda \to -\hrho,\qquad \hrho\to
2\hrho + \hlambda,\qquad \hsigma\to \hsigma+2\hrho,
}
while   $S$ is given by
\eqn\eSTrans{
  \lambda \to \hsigma+2\hrho,\qquad \hrho\to -\hrho ,\qquad \hsigma\to
  \hlambda+2\hrho.}

The divergences of the loop amplitude
are particularly easy to classify in terms of
integration over  $\tau$ and $V$  \eintom.
The leading and subleading divergences arise from
 two distinct kinds of boundaries of the integration
domain.
\hfill\break\indent
  $(I)$:\ \
The leading  divergence arises  from the limit $V\to \infty$ with arbitrary fixed
values
of $\tau_1$ and $\tau_2$.  We will regulate this by cutting off the upper limit
at a value $V=V^c = a\, \Lambda^2 l_{11}^2$ (where $a$ is an arbitrary constant)
so that the amplitude is proportional to
$\Lambda^8$.   This is  the two-loop primitive divergence
which comes from  the region in which the loop momenta  are  simultaneously of
order $\Lambda$ and  corresponds to the region in which all three
Schwinger parameters approach their lower end-points.
\hfill\break\indent
 $(II)$:\ \
 The three distinct kinds of
 subleading ultraviolet divergences  arise  from the region in which $\tau_2 \to
 \infty$ with $V$ fixed, together with the $SL(2,{\bf Z})$ images of this region obtained
 by the action of $S$ and $TS$.
 These are the divergences which arise when  one of the winding parameters
 approaches its upper limit, which is cut off at $\Lambda$.  From \modef\
this translates into a cutoff on the upper $\tau_2$ limit at $\tau_2^c=
\Lambda^2 l_{11}^2 V^{-1}$, which  means that
the  complex structure is integrated over the
restricted fundamental domain,
\eqn\resft{ \calF_{\tau_2^c}=\{ -\half \leq \tau_1 \leq \half, \; \tau_2\leq
  \tau_2^c, \; \tau_1^2+\tau_2^2\geq 1\}.}
It is easy to see that this includes all the sub-divergences, as follows. When
 $\tau_2$ reaches its upper limit with fixed $V$,
 both $\hrho$ and $\hlambda$ approach zero while $\hsigma$ becomes infinite.  This
 translates, via \defsn, into the region in which $\rho$ and $\lambda$
 vanish, which corresponds to the ultraviolet
 sub-divergence at which   $q^2\to \infty$.
Similarly, the image under $S$ of this limit is the boundary at $\tau_1=0,
\tau_2=0$ which corresponds to the sub-divergence at which $(p+q)^2\to \infty$.  The
image under $TS$ is the boundary at $\tau_1=1, \tau_2=0$ which  corresponds
to the sub-divergence at which  $p^2\to \infty$.

 We will now see how this description of the divergences is particularly
 well adapted to  compactification on a toroidal target space since cases (I) and (II)
 arise from distinct classes of $SL(2,\ZZ)$ orbits in
  the mapping of $\hat \calT^2$ into $\calT^2$.

%%%%%%%%%%%%%%%%%%%%%%%%%%%%%%%%%%%%%%%%%%%%%%%%%%%%%%%%%%%%%%%%%%%%%%%%%%%%%%
\subsec{Compactification on $\calT^2$}

When  the eleven-dimensional two-loop amplitude is compactified on a
  two-torus ($n=2$) of volume $\calV$ and complex structure $\Omega$
   the exponential factor
  \eEw\ can be written as
\eqn\eewt{
E_w = {\calV V\over \Omega_2 \tau_2} \left|(1\ \Omega)  A  (\tau\
  1)\right|^2 - 2\calV V \det(A)\ ,
}
where we have used the usual formula for the metric on a two-torus,
\eqn\usug{
G_{IJ} \hm_I \hm_J =l_{11}^2\calV {|\hm_1+\hm_2\Omega|^2\over \Omega_2},}
and defined  a $2\times 2$ matrix A
%\eqn\adef{A=\pmatrix{k&m\cr l&n\cr},}
with integer entries.

In this case the expression \eintom\ becomes
\eqn\eun{
\eqalign{
I^{P+NP}(0)&={2\pi^7\over l_{11}^8}  \sum_{{\hm_I},{\hn_I}}
\int_0^{V^c} dV V^3 \int_{\calF_{\tau_2^c}}
{d^2\tau\over \tau_2^2} \exp\left(-\pi{\calV \over \Omega_2}{V\over
    \tau_2} \left|(1\ \Omega)  A  (\tau\ 1)\right|^2 + 2\pi\calV V
  \det(A)\right)\cr
&={2 \pi^7\over l_{11}^8} \int_0^{V^c}
 dV V^3 \int_{\C|F_{\tau_2^c}}
{d^2\tau\over \tau_2^2} \tau_2 \Gamma_{(2,2)}(\calV,\Omega; V, \tau).
}}
This expression can be analyzed by splitting it into orbits for the right action
of $Sl(2,\ZZ)$ on $\tau$.
There are three classes of orbits:
\item{(I)}
Singular orbits are ones with $A=0$.  By inspection, it is clear that these
give the leading divergent contributions   which are
proportional to $\Lambda^8$ from the boundary $V=V^c =a\, l_{11}^2\Lambda^2$.
\item{(II)} Degenerate orbits are those for which  $\det A=0$.  In
this case $A$ can be transformed to the form $A=\pmatrix{0&l\cr
0&k}$ by a $SL(2,\ZZ)$ transformation that maps the fundamental domain to the strip.  Again, by inspection these can be seen
to give the
sub-divergences proportional to $\Lambda^3$.
\item{(III)} Non-degenerate orbits are ones with non-singular $A$.  An
$SL(2,\ZZ)$ transformation, that maps the fundamental domain to the upper-half
complex plane, can be used to put $A$ in the form
 $A=\pm \pmatrix{m&j\cr0&n}$ where  $0\le j \le m-1$ , $m>0$  and $n\neq0$
\refs{\rfDixonKapluLouis,\rfBFKOV}.  The non-degenerate orbits
are the ones that contribute to the finite part of the two-loop
integral after the divergent terms have been subtracted out.
Since the string coupling constant appears in $\Omega$ both the
degenerate and the non degenerate cases will contribute to perturbative and
non-perturbative string effects.

\noindent
The singular term (I)
 comes from the zero winding number sector and does not depend on the shape of the
torus.  A term of the same form also arises by adding a local two-loop counterterm
proportional to $S^2\, \calR^4$ to the effective action.
We will argue later   that consistency
with string perturbation theory actually requires
 this renormalized  coefficient of this
 interaction to  vanish.

In order to extract the remaining contributions
it is useful to first consider the action of the  Laplace
operator $\Delta_\Omega =4\Omega_2^2\, \partial_\Omega \bar\partial_\Omega$
on the expression \eun\   using the identity,
\eqn\eLaplace{
\Delta_\Omega \left( \tau_2 \Gamma_{(2,2)} \right)
 =  4\tau_2^2 \partial_\tau \bar\partial_\tau
 \left( \tau_2 \Gamma_{(2,2)} \right) =\Delta_\tau \left(
 \tau_2 \Gamma_{(2,2)} \right) .}
It follows that
\eqn\ediffi{\eqalign{
\Delta_\Omega I(\calV,\Omega ) = &
\quad {2\pi^7\over l_{11}^8}
\int_0^{V^c}
 dV V^3 \int_{\calF_{\tau_2^c}} d^2\tau
\cr &
(\partial^2_{\tau_1}+ \partial^2_{\tau_2})
 \sum_{\{m^I,n^I\}} \exp\left(-\pi{\calV \over \Omega_2}{V\over
      \tau_2} \left|(1\ \Omega) A (\tau\ 1)\right|^2 + 2\pi\calV V
    \det(A)\right) ,\cr}
}
so that the $\tau$ integration at fixed $V$ is simply the integral of a
surface term which gets contributions from the boundary of the integration
domain $\calF_{\tau_2^c}$ at
$\tau_2 = \tau_2^c \equiv l_{11}^2\Lambda^2/V$. The expression
\ediffi\ reduces to
\eqn\eboundi{
\eqalign{
\Delta_\Omega I(\calV,\Omega ) &= {2\pi^7\over l_{11}^8}
\int_0^{V^c}\!\!\!\! dV V^3
\,\partial_{\tau_2} \!\!\!\!\!\left.\sum_{A\in M(2,2,\ZZ)}\!\!\!\!
\exp\left(-\pi{\calV \over \Omega_2}{V\over \tau_2} \left|(1\ \Omega)  A
(\tau\ 1)\right|^2 + 2\pi \calV V \det(A)\right)
\right|_{\tau_2=\tau_2^c}\cr
&=-{ 2\pi^8\calV\over l_{11}^8} \int_0^{V^c}
 dV V^4 \!\! \sum_{A\in M(2,2,\ZZ)}
\partial_{\tau_2}\left( \left|(1\
      \Omega)  A (\tau\ 1)\right|^2\over \tau_2\Omega_2\right)\times \cr
&\kern 4cm
\left.\exp\left(-\pi{\calV \over \Omega_2}{V\over \tau^c_2} \left|(1\
\Omega)  A (\tau^c\ 1)\right|^2 + 2\pi\calV V
\det(A)\right)\right|_{\tau_2=\tau^c_2}.
}}
In the  limit $\Lambda^2\to\infty$  all the terms in the sum are exponentially
suppressed apart from the degenerate orbits $A=\pmatrix{0&l\cr 0&k}$.  In this
sector the exponent evaluated at the boundary
$\tau_2 =\tau_2^c= \Lambda^2 l_{11}^2/V$  reduces to
\eqn\expnon{E_w = {\calV V\over \Omega_2 \tau^c_2} |l+k\Omega|^2 =
 \calV {|l+k\Omega|^2\over \Omega_2} {V^2\over l_{11}^2\Lambda^2},}
so that  \eboundi\ reduces to
\eqn\eboundj{
\eqalign{
\Delta_\Omega I(\calV,\Omega )&= {2\pi^8 \calV\over l_{11}^{12}\Lambda^4}
\int_0^\infty dV V^6 \sum_{(l,k)\neq (0,0)} {|l+k \Omega |^2\over \Omega_2}
e^{-\pi\calV {|l+k\Omega|^2\over \Omega_2} {V^2\over l_{11}^2\Lambda^2}}\cr
&= 2\pi^{9\over2} \Gamma({\textstyle{7\over2}})\zeta(5) \Lambda^3\; l_{11}^{-5}\calV^{-{5\over
    2}}E_{5\over 2}(\Omega),
}}
where the upper limit is taken to infinity since the integral converges.
Since $E_{5/2}(\Omega)$ satisfies the Laplace equation
\eqn\lapeq{
\Delta_\Omega E_{5\over2}(\Omega)={15\over 4} E_{5\over2}(\Omega),
}
we conclude that the two-loop integral has the form
\eqn\genres{
I(\calV,\Omega) =a\,
\Lambda^8+ \pi^5\zeta(5) \Lambda^3\; l_{11}^{-5} \calV^{-{5\over 2}} E_{5\over
  2}(\Omega)
+ I_{fin}.}
The first term, which has an undetermined value, is the leading regularized
divergence and arises from the singular orbits. Its coefficient
 is modified by the addition
of the two-loop $S^2\, \calR^4$ counterterm with coefficient $c_2$.  The second
term in \genres\  is the
contribution of the degenerate orbits. This has a $\Lambda$-dependent
coefficient to which must be added the contribution that comes from
  \ftwo(b) which includes the effect of the one-loop counterterm.  As will be seen in  section 4.3 the combined coefficient is  consistent with  the   equality of the IIA and IIB string theory one-loop and two-loop amplitudes.

 The term $I_{fin}$ in \genres\  is independent of
the cutoff and is the finite term that comes from the non-degenerate orbits
and  must satisfy $\Delta_\Omega \,  I_{fin}=0$ . It can be evaluated
explicitly using the `unfolding trick'.  This allows one of the infinite sums
in \eun\ to be used to rewrite the $\tau$ integral over $\calF$ as an integral
over the upper-half $\tau$ plane.  This is similar to the analysis in
\refs{\rfBFKOV}, from which the result can be extracted in the form
\eqn\enkind{
I_{fin} = {4\pi^3\over l_{11}^8} \sum_{m>0,n\neq 0\atop 0\leq j<m} \int_0^\infty
 dV V^3 \int_{\CC^+} {d^2\tau\over \tau_2^2} \, e^{-{\calV V\over \Omega_2
    \tau_2}|m\tau + (j + n\Omega)|^2 + 2\calV Vmn} .}
In this case no cutoff is necessary and the result has a unique normalization.
The  $\tau_1$ integration is gaussian and can be carried out
explicitly giving,
\eqn\omhin{
I_{fin} ={4\pi^{7\over2}\over l_{11}^8}\sqrt{\Omega_2\over \calV} \sum_{0\leq
  j<m\atop m>0,n\neq 0} {1\over m}  \int_0^\infty dV V^3 \int_0^\infty
{d\tau_2\over \tau_2^2} \sqrt{\tau_2\over V} e^{2 V \calV mn} e^{- {\calV\over
\Omega_2} {V\over \tau_2} (m\tau_2 + n\Omega_2)^2}.}
Now setting $x=V/\tau_2$ and $y=V\tau_2$ we have
\eqn\etornon{
\eqalign{
I_{fin} &= {2 \pi^{7\over2}\over l_{11}^8}\sqrt{\Omega_2\over \calV} \sum_{0\leq j<m\atop
m>0, n\neq 0} {1\over m} \int_0^\infty dx\; x\int_0^\infty dy\; y^{1/2}\
e^{-{l_{11}^2\calV\over \Omega_2} (m^2y + n^2\Omega_2^2 x )}\cr
&=  2\pi^4 \zeta(3)\zeta(4)\; l_{11}^{-8} \calV^{-4}\ .\cr
}}

%%%%%%%%%%%%%%%%%%%%%%%%%%%%%%5
\subsec{Contribution from one-loop and two-loop counterterms}

The sum of the contributions to the  amplitude  from
 \ftwo(b) and \ftwo(c) will be denoted  $\delta  A_4^{(2)}
= \delta_1 A_4^{(2)}  + \delta_2 A_4^{(2)}$.
The term  $\delta_1 A_4^{(2)}$  corresponds to \ftwo(b) and
is proportional to the one-loop counterterm so it has an overall factor of
$c_1$, which has the value given by \regcon.  The direct evaluation of
this process would  require a complicated sum over the different
types of particles circulating in the loop. However,
 it is easy to check that the prescription of
\refs{\rfBernDunbar} for expressing the one-loop and two-loop
supergravity diagrams
in terms of scalar field theory Feynman rules generalizes to
diagrams of this type,  giving,
\eqn\corram{\delta_1 A_4^{(2)}=\! ic_1{\pi^3\kappa_{11}^6\over
2(2\pi)^{22}\, l_{11}^3}\;
 \hK\!
(S^2 + T^2+U^2)\, \delta_1 I,}
where the loop integral $\delta_1 I$ is given by using scalar field propagators and
vertices in \ftwo(b),
\eqn\variam{\delta_1 I = \int d^{11}q\,
 {1\over q^2} {1\over (q+k_1)^2}{1\over (q+k_1+k_2)^2}.}
The normalization in \corram\ can be obtained as  a simple consequence
of  unitarity.

The evaluation of the integral \variam\ compactified on $\calT^2$
 follows closely the   discussion in section 3 of the box diagram.
 The only difference is that in this case there are only
 three internal propagators.  The result is
\eqn\rescon{\delta_1 I  =  {\pi^{1\over 2}\over l_{11}^5}
\left( (\Lambda l_{11})^5  + \calV^{-{5\over2}}
2\zeta(5)\, \Gamma({\textstyle{5\over2}})\, E_{5\over 2}(\Omega,\bar \Omega)\right).}
The cutoff-dependent term comes from the zero winding sector,
and  upon inserting \rescon\ in \corram,  contributes a term proportional to
$\Lambda^8$ to the
 $\C|V$-independent part of  the amplitude.   Its coefficient will be
 absorbed into a redefinition of the coefficient  $a$
  of the leading two-loop divergence 
 in \genres.
The $\Omega$-dependent part of \rescon\ has the same form as the
$\Lambda^3$ sub-divergences of
the two-loop amplitude in \genres.  After adding these two contributions
and  substituting the value of $c_1$ from \regcon\ the net
dependence on the cutoff cancels, leaving a specific finite
contribution that will be discussed in the following subsection.

The  contribution to $\delta_2 A_4^{(2)}$ from the two-loop local
counterterm is  equal to
\eqn\secdef{\delta_2 A_4^{(2)} =i c_2
 {\kappa_{11}^6\over(2\pi)^{22}l_{11}^8}\, \hK\,
(S^2+T^2+U^2),} where $c_2$ is a constant which, for the moment, is arbitrary.

%%%%%%%%%%%%%%%%%%%%%%%%%%%%%%%%%%%%%%%%%%%%%%%%%%%%%%%%%%%%%%%%%%%
\subsec{Comparison of eleven-dimensional  supergravity and type II
string theories.}

We now turn to the comparison of the results of the
eleven-dimensional calculations to those of the  type II string
theories.  We will check that the normalization of the finite
$S^2\,\calR^4$
term \etornon\ has precisely the value that is needed for the
perturbative type IIA and IIB string theories to be equal at the order of
one string loop.  Furthermore, the value of the
one-loop counterterm, \regcon,  will also be  seen to lead to the
equality of the
IIA and IIB string tree-level and two-loop
terms.  This strongly supports the impression that the
two-loop contribution to $S^2\,\calR^4$ does not get further
contributions from higher-order Feynman diagrams.

In order to compare our two-loop  supergravity results with string theory it is
 necessary to carefully  specify our conventions.
In either of the two string theories the   four-graviton ampitude
 has the expansion at tree-level and one loop,
\eqn\esexp{
A_4^{string}=  \kappa_{10}^2\hK \, \left[- e^{-2\phi} T +
  {\kappa_{10}^2\over 2^5 \pi^6l_s^8}   I^{1-loop} + \cdots \right],}
where the terms in the square brackets are dimensionless (recall that
$2\kappa_{10}^2 =(2\pi)^7 l_s^8$)  as in the analysis
of~\refs{\rfGreenVanhoveAnalytic}.  The low-energy expansion of the
tree-level and one-loop terms, $T$ and $I^{1-loop}$,
 are briefly described in the appendix and in \refs{\rfGreenVanhoveAnalytic}.
The one-loop and two-loop amplitudes
in eleven-dimensional supergravity, together with
the effects of the counterterms,
  are given  by
\eqn\emexp{\eqalign{
A_4 + \delta A_4 = & {\kappa_{11}^4\over  (2\pi)^{11}}  \hK \left[
I(S,T)+I(S,U)+I(T,U)\right]  \cr
 &+i{\kappa_{11}^6\over (2\pi)^{22}}  \, \hK \, \left[ S^2 (I^P(S,T) +
I^{NP}(S,T)+I^P(S,U) + I^{NP}(S,U) ) + perms.\right] \cr
&  +\delta A_4^{(1)}+
\delta_1 A_4^{(2)}+ \delta_2 A_4^{(2)} .\cr}}

The results of section 3 show that  the expansion up to  order
$S^2$ of the one-loop supergravity amplitude compactified on $\calT^2$
is given by
\eqn\emone{\eqalign{
A_4^{(1)} + \delta A_4^{(1)}
= & {\kappa_{11}^4\over (2\pi)^{11}l_{11}^3} \,
\hK \left[ {2\zeta(3)\over R_{11}^3}+{4\zeta(2)\over
R_{11}R_{10}^2} +  {2\pi^2\over 3} \right. \cr
& \left. \quad + l_{11}^4 (S^2+T^2+U^2) {\pi^2\over 6!} \left(4\zeta(2) R_{11}
+ 2\zeta(3) {R_{10}^2\over R_{11}} \right) + \cdots  \right],\cr
}}
where  $\cdots$
indicates the infinite series of D-instanton contributions
\refs{\rfGreenGutperleVanhove}. Converting  into type IIA  string variables this
becomes
\eqn\emsone{
\eqalign{
A_4^{(1)} + \delta A_4^{(1)}  =& (4\pi^8 l_{11}^{15}r_A^{-1})r_A    \hK \left[
  2\zeta(3)e^{-2\phi^A} +{4\zeta(2)\over r_A^2}+
 {2\pi^2\over 3}\right. \cr & \left. \quad + l_s^4 (s^2+t^2+u^2)
  {\pi^2\over 6!} \left(4\zeta(2) e^{2\phi^A} + 2\zeta(3) r_A^2 \right) +
  \cdots \right].
}}

The two-loop result can be written as
\eqn\emtwo{
\eqalign{
A_4^{(2)} + \delta A_4^{(2)}  = & i{\kappa_{11}^6\over(2\pi)^{22}l_{11}^8} \times 
\hK (S^2+T^2+U^2) \left[a(\Lambda l_{11})^8 + c_2\right.\cr
&\left.\quad  +  {\pi^6\over 4}\left({2\zeta(5)\over
      R_{11}^5} + {8\over 3} \zeta(4) {1\over R_{11} R_{10}^4} \right) +
  {2\pi^4\zeta(3)\zeta(4)\over (R_{10}R_{11})^4} +\cdots \right],
}}
where we have expanded the modular function $E_{{5\over
2}}(\Omega,\bar\Omega)$ in powers of $\Omega_2^{-1} = R_{11}/R_{10}$.
The constant $a$ is meant to represent the sum of the primitive
two-loop divergences  that arise from the zero
winding number sectors  of \fthree\ and \ftwo(b).
These combine with the coefficient of the two-loop counterterm, $c_2$.
After conversion into  type IIA string variables this becomes
\eqn\emstwo{\eqalign{
A_4^{(2)} + \delta A_4^{(1)} =&(4\pi^8 l_{11}^{15}r_A^{-1}){i\over8\pi^6}
r_A\hK l_s^4(s^2+t^2+u^2) \left[(a(\Lambda l_{11})^8+c_2)
e^{4\phi^A/3}\right. \cr
&\left. \quad +  {\pi^6\over 4}\left(2\zeta(5)e^{-2\phi^A} + {8\over 3}
\zeta(4) {e^{2\phi^A}\over
      r_A^4} \right)  + {2\pi^4\zeta(3)\zeta(4)\over r_A^4}+\cdots  \right].}}

Now we can use  T-duality to replace the last term in \emstwo, which has the
dilaton dependence of a string one-loop term, by its IIB
equivalent, which is proportional to $r_B^2$ and is also identified as a
one-loop string contribution.  Using the fact that the two string theories
have identical  four-graviton loop amplitudes (up to two loops) we must
identify this term with the term in parentheses in  \emsone\
that is  proportional to $r_A^2$ (which
was deduced from a one-loop effect in eleven-dimensional
supergravity).  Gratifyingly the coefficients of these terms are
indeed equal (using $\zeta(4) = \pi^4/90$) which appears to be a
rather nontrivial check on our calculation.
Similarly, we can  check that the renormalized value for the subleading
divergences  respects this symmetry between the IIA and IIB theories
since  the term   proportional to $e^{2\phi^A}$ in \emsone\ has the
same coefficient
as the term proportional to $e^{2\phi^A}$ in \emstwo.
Now we can check the consistency further  by comparing the
coeffcient of the
tree-level term   proportional to $\zeta(5) e^{-2\phi^A} s^2\calR^4$
in
\emstwo\  with the coefficient of the
tree-level term proportional  to $\zeta(3)e^{-2\phi^A} \calR^4$ in
\emsone.  These  coefficients
 agree  with the coefficients deduced from the expansion of the
four-graviton tree amplitude reviewed in the appendix.

We also need to consider the value of the leading divergent
contribution to the $S^2\, \calR^4$ interaction which  arises from
the loop amplitudes combined with the two-loop
counterterm with coefficient $c_2$
 and is independent of the parameters
of the two-torus. Translating into  IIA string theory coordinates this gives the term
proportional to  $R_{11}^2 \, s^2\,\calR^4= e^{4\phi^A/3}\,
s^2\,\calR^4$ in \emstwo\  which
is not proportional to an integer power of $e^{2\phi^A}$.  Therefore,  it
cannot possibly arise in string perturbation theory, which means that its
coefficient must vanish, i.e.
\eqn\ctwop{c_2+a(\Lambda l_{11})^8=0.}
 A consequence of
this is that the
$D^4\, \calR^4$ interaction vanishes in the decompactification limit,
$R_{11}\to \infty$ so there is no $D^4\calR^4$ interaction in the
eleven-dimensional theory.

%%%%%%%%%%%%%%%%%%%%%%%%%%%%%%%%%%%%%%%%%%%%%%%%%%%%%%%%%%%%%%%%%%%%%%%%%%%
\newsec{Conclusion}

In earlier work  the
$\calR^4$ interaction in the effective action for eleven-dimensional M
theory compactified on $\calT^2$
 was obtained by evaluating one-loop Feynman diagrams.
Although the dependence on the complex structure was uniquely
determined by this supergravity calculation,
in order to pin down the value of the one-loop
counterterm  it was necessary to input the extra information that
the  four-graviton amplitudes in  type IIA and type IIB superstring
theories are equal at one string loop.

In this paper  we have generalized these statements to obtain
the exact scalar field dependence  of the coefficient of the
$S^2\, \calR^4$ interaction based on consideration of two-loop Feynman
diagrams for four-graviton scattering in  eleven-dimensional
supergravity.  Using the value of the one-loop counterterm determined from
the one-loop analysis, we have found that the renormalized  value of the
two-loop amplitude is
\eqn\fullres{\eqalign{
\C|S_{\C|D^4\C|R^4}=&  {l_{11}^3\over 48\cdot (4\pi)^7}\, \int d^9x\,
\sqrt{-G}\, \C|V\,D^4\calR^4  \left(   \zeta(5)  \C|V^{-{5\over2}}
  E_{5\over 2}(\Omega,\bar\Omega)+ {4\over \pi^2} \zeta(3)\zeta(4)\,
\calV^{-4} \right) \cr
= &  {l_s^3\over 48\cdot (4\pi)^7} \int d^9x \sqrt{-g^B}\; r_B\;
D^4\C|R^4
\left( \zeta(5) e^{\half\phi^B}  E_{5\over 2}(\Omega,\bar\Omega)
+  {4\over\pi^2} \zeta(3)\zeta(4)\; r_B^2 \right) \,  ,\cr}}
where the second equality expresses the amplitude in type IIB
parameters, recalling from  \defcals\ that  $D^4 \calR^4$ is a symbolic way of
representing the contraction of covariant derivatives and curvature
tensors that gives rise to the kinematic factor $(S^2+T^2+U^2)\hK$ in the
four-graviton scattering amplitude.

The term in parentheses in \fullres\ that is independent of
$\Omega$ matches a corresponding term that arises in the type IIA
parameterization
from the one-loop supergravity amplitude in \twonterm. It should be easy to
evaluate the string one-loop amplitude in nine dimensions and check
the coefficient of this term. We saw in
section 3 that terms of this type, which appear to be singular in the
decompactification limit, $r_B \to \infty$, sum up to form the
appropriate massless threshold singularity in ten dimensions.
  The dependence on the complex structure of
the torus (the scalar field of the IIB theory) is  contained entirely in the
modular function   $E_{5/ 2}(\Omega,\bar \Omega)$ which survives the
decompactification limit to the ten-dimensional type IIB theory.  This
term  has an expansion in the
coupling $e^{\phi^B}$ that begins with a  tree-level term followed by a
two-loop term and then an infinite series of D-instanton contributions. We
have seen that this is  consistent with the little that  is known from string
perturbation theory  -- the tree-level coefficient agrees with the string tree
calculation reviewed in the appendix and the one string loop contribution to
the $s^2\, \calR^4$ interaction in ten dimensions is absent as it should be
according to \refs{\rfGreenVanhoveAnalytic}.  However,
since no two-loop string amplitudes have yet been evaluated, the value we have
obtained  for the $s^2\, \calR^4$ interaction at two string loops is not yet
tested (although precise two-loop string calculations are feasible in
principle  \refs{\rfIengoZhu}).  The same is true for the infinite sequence of
D-instanton contributions to this interaction.

Although we saw in the last section that the  $S^2\, \calR^4$ term
cannot contribute to  the eleven-dimensional theory   in the decompactification limit, $\calV \to \infty$, the  next
 term in the derivative expansion of the two-loop amplitude may.  This is the
 term  of the
 form  $S^3\, \calR^4$ which translates in type  IIA string parameters
to a term of
 the form $e^{2\phi^A}\, s^3\, \calR^4$ which would be a  string two-loop
 effect.   When $\calV$ is finite there are  other supergravity two-loop
contributions to the prefactor multiplying
$S^3\,   \calR^4$ which come from the expansion of the planar and
nonplanar diagrams,
 \enonpll\ and \enonnon, to linear order in $S$, $T$ and $U$.  The resulting
 expressions do not possess modular invariant integrands when expressed in
 terms of the integration variables $V$ and $\tau$ and we have not made sense of the
 integrals.   This suggests that extra
 contributions from higher-loop supergravity amplitudes
 are needed to give the full  form of the
 prefactor.  Since there are good dimensional
 arguments to expect this  prefactor  to be determined by supersymmetry it would
 be of interest to disentangle these contributions.

%%%%%%%%%%%%%%%%%%%%%%%%%%%%%%%%%%%%%%%%%%%%%%%%%%%%%%%%%%%%%%%%%%%
\vskip 0.5cm
{\it Acknowledgements:}  We are grateful to Zvi Bern and Lance Dixon for
interesting discussions and to  the  Theory Division at CERN where part of
this work was accomplished. P.V. is also grateful to the Service de Physique
th{\'e}orique de Saclay and the Laboratoire de physique th{\'e}orique
de l'{\'E}cole
normale sup{\'e}rieure de Paris for hospitality as well as to  PPARC
for financial  support.

%%%%%%%%%%%%%%%%%%%%%%%%%%%%%%%%%%%%%%%%%%%%%%%%%%%%%%%%%%%%%%%%%%%%%%%%%%%%
\appendix{A}{The structure of the tree-level and one-loop string
  contributions}

The sum of the tree-level and one-loop contributions to the four-graviton
amplitude  in ten dimensions has the form
\refs{\rfGreenSchwarz,\rfGreenSchwarzWitten,\rfDhokerPhongRevue},
\eqn\etreeone{
A^{string}_4 =\kappa_{10}^2 \hK
\left[-e^{-2\phi}\, T(s,t,u) + {\kappa_{10}^2\over 2^5
\pi^6l_s^8 } \int_\C|F {d^2\Omega\over \Omega_2^2} F(\Omega,\bar \Omega;s,t,u)
\right],
}
where the functions $T$ and $F$ contain the dependence on the Mandelstam
invariants of the tree-level and one-loop terms, respectively,
$2\kappa_{10}^2= (2\pi)^7  l_s^8 $ is defined as in
\refs{\rfGreenVanhoveMtheory} and $d^2\Omega=d\Omega_1 d\Omega_2$.

The function $T$ contains the dynamical part of the
 tree amplitude for the elastic scattering of two gravitons
in either type~II  theory and  is given
\eqn\eTree{
\eqalign{
T &= {64\over l_s^6  stu}
{\Gamma(1-{l_s^2\over4} s)\Gamma(1-{l_s^2\over 4} t)\Gamma(1-{l_s^2\over 4} u)
 \over \Gamma(1+ {l_s^2\over 4} s)\Gamma(1 + {l_s^2\over 4} t)\Gamma(1  +
 {l_s^2\over 4} u)} \cr
&= {64 \over l_s^6 stu} \exp\left(\sum_{n=1}^\infty {2 \zeta(2n+1) \over
 2n+1}{l_s^{4n+2}\over 4^{2n+1}} (s^{2n+1} + t^{2n+1} + u^{2n+1})\right)\ .
}}
Thus, the low energy expansion of the amplitude begins with the
terms,
\eqn\etreeexp{
\eqalign{  T = {64\over l_s^6 stu} & +
2\zeta(3)   +  {\zeta(5)\over16}l_s^4 (s^2+t^2+u^2)\cr
& + {\zeta(3)^2\over 96}  l_s^6 (s^3+t^3+u^3) +{\zeta(7)\over 512}
 l_s^8 (s^2+t^2+u^2)^2 + \dots \ .\cr
}}

This can be rewritten in terms of the coordinates of the
eleven-dimensional theory by using the dictionary \usdico.
When expressed in terms of the Mandelstam invariants in the M-theory metric
the  expression \etreeexp\  has the low-energy expansion,
\eqn\emexp{ \eqalign{T\; R_{11}^{-3}=  {64\over l_{11}^6 STU} & +
{2\zeta(3)\over R_{11}^3}   + {\zeta(5)\over 16R_{11}^5} l_{11}^4
(S^2+T^2+U^2) \cr
&+  {\zeta(3)^2\over 96 R_{11}^6}
l_{11}^6(S^3+T^3+U^3)+{\zeta(7)\over 512 R_{11}^7}l_{11}^8(S^2 + T^2 + U^2)^2
}}
The dynamical factor $F$ in the loop amplitude is given in
terms of the scalar Green function on the torus $\ln \chi_{ij}$,
\eqn\eF{
F(\Omega,\bar \Omega; s,t,u) =  \int_{\calT^2} {d^2\Omega\over
\Omega_2^2}\prod_{i=1}^3
{d^2\nu^{(i)}\over \Omega_2}
\left(\chi_{12}\chi_{34}\right)^{l_s^2 s} \left(\chi_{14}\chi_{23}
\right)^{l_s^2 t} \left(\chi_{13}\chi_{24} \right)^{l_s^2u}\ .
}
The  low-energy expansion of this expression is  considered in
\refs{\rfGreenVanhoveAnalytic} where the first few terms are
shown to be,
\eqn\eoneexp{
\eqalign{
A^{one-loop}_4 = 2\pi\kappa_{(10)}^2\,\hK\left( {\pi\over 3}+  {l_s^2\over 16} I_{nonan\, 1} +\right.&
0\times l_s^4(s^2+t^2+u^2) \cr
&\left.+{\pi\over 36} \zeta(3)l_s^6(s^3+t^3+u^3) + {l_s^8\over 256}
  I_{nonan\, 2} + \cdots \right)\cr
  = \hK\left({\pi\over 3}+ {l_{11}^2\over 16} I_{nonan\, 1} +\right.& 0\times l_{11}^2
(S^2+T^2+U^2)\cr
&\left.+{\pi\over 36}\zeta(3) l_{11}^6 (S^3+T^3+U^3)+{l_{11}^4\over
256} I_{nonan\, 2}  + \cdots
\right).
}}
The functions  $I_{nonan\, 1}$ and $I_{nonan\, 2}$ are
non-analytic terms that contain logarithmic contributions to the
two-particle normal thresholds that are defined by $I'$ in \iprimedef\
and are given more explicitly in
\refs{\rfGreenVanhoveAnalytic}.

\listrefs

\bye